\documentclass[3p,twocolumn]{elsarticle}
\biboptions{authoryear}

\usepackage[T1]{fontenc}
\usepackage{ae,aecompl}

\usepackage{hyperref}
\usepackage{graphicx}	
\usepackage{amsmath}	
\usepackage{amssymb}	
\usepackage{epstopdf}
\usepackage{color}
\usepackage{hyperref}

\newcommand{\cor}[1]{}
\newcommand{\sub}[1]{_{\rm #1}}
\renewcommand{\sup}[1]{^{\rm #1}}

\newcommand{\MEarth}{M\sub{\oplus}}
\newcommand{\REarth}{R\sub{\oplus}}
\newcommand{\Pb}{Proxima b}
\newcommand{\Pba}{\Pb\ analogues}
\newcommand{\pa}{planetary analogues}
\newcommand{\na}{nominal analogues}
\newcommand{\gp}{magnetic properties}
\newcommand{\iac}{in all cases}

\newcommand{\Mdip}{{\cal M}\sub{dip}}
\newcommand{\MdipEarth}{{\cal M}\sub{dip,\oplus}}
\newcommand{\dtot}[2]{ \frac{ d #1 }{d #2} }
\newcommand{\beq}[1]{\begin{equation}\label{#1}}
\newcommand{\eeq}{\end{equation}}
\newcommand{\pr}[1]{ \left( #1 \right) }
\newcommand{\der}[2]{ \frac{ \partial #1 }{\partial #2} }
\newcommand{\hl}[1]{{\color{black} #1}}
\newcommand{\hll}[1]{{\color{black} #1}}
\newcommand{\hlll}[1]{{\color{black} #1}}

\def\apj{ApJ}

\usepackage{hyperref}

\journal{Planetary and Space Sciences}

\begin{document}

\begin{frontmatter}

\title{Magnetic properties of Proxima Centauri b analogues}

\author[udea]{Jorge I. Zuluaga\corref{corr}}
\cortext[corr]{Corresponding author}
\ead{jorge.zuluaga@udea.edu.co}

\author[heilderberg]{Sebastian Bustamante}
\ead{sebastian.bustamante@h-its.org}

\address[udea]{Solar, Earth and Planetary Physics (SEAP), FACom\\
Instituto de F\'isica-FCEN, Universidad de Antioquia, Cl 70
No. 52-21, Medell\'in, Colombia}

\address[heilderberg]{Heidelberg Institute f\"ur Theoretische Studien \\ Schloss-Wolfsbrunnenweg 35, 69118 Heidelberg, Germany}

\begin{abstract}
The discovery of a planet around the closest star \hlll{to our Sun}, Proxima Centauri, represents a quantum leap in the testability of exoplanetary models.  Unlike any other discovered exoplanet, models of Proxima b could be contrasted against near future telescopic observations and far future in-situ measurements. In this paper \hlll{we aim at predicting} \hl{the planetary radius} and the magnetic properties (dynamo lifetime and magnetic dipole moment) of Proxima b analogues \hl{(solid planets with masses of \hlll{$\sim 1-3\,M_\oplus$}, rotation periods of several days and habitable conditions). For this purpose we build a grid of planetary models with a wide range of compositions and masses.  For each point in the grid we run the planetary evolution model developed in \citet{Zuluaga2013}.  Our model assumes small orbital eccentricity, negligible tidal heating and earth-like radiogenic mantle elements abundances. We devise a statistical methodology to estimate the posterior distribution of the desired planetary properties assuming simple \hll{prior distributions} for the orbital inclination and bulk composition. Our model predicts that Proxima b \hlll{would} have a mass $1.3\leq M_{\rm p}\leq 2.3\,M_{\oplus}$ and a radius $R_{\rm p}=1.4^{+0.3}_{-0.2}\,R_{\oplus}$.  In our simulations, most Proxima b analogues develop intrinsic dynamos that last for $\geq$4 Gyr (the estimated age of the host star).  If alive, the dynamo of \Pb\ have a dipole moment ${\cal M}_{\rm dip}>0.32^{\times 2.3}_{\div 2.9}{\cal M}_{\rm dip,\oplus}$.  These results are not restricted to \Pb\ but \hlll{they also} apply to earth-like planets having similar observed properties.}
\end{abstract}


\end{frontmatter}



\section{Introduction}

After 16 year of observations, a planet around the closest star \hlll{to our Sun}, Proxima Centauri or Proxima, has been discovered \citep{AngladaEscude2016}. The reported properties of the planet, a minimum mass of 1.3$\MEarth$ and orbital period about $11.3$ days, which places it inside the habitable zone (HZ), make \Pb\ the closest Earth-twin. Beyond its obvious astrobiological relevance \citep{Barnes2016,Ribas2016,Turbet2016}, future observations of \Pb\ with ground- and space-based telescopes (E-ELT, JWST, WFIRST, etc.) and, hopefully, in-situ space probes \citep{Lubin2016}, will provide us contrasting information for models of planetary processes not yet tested beyond the Solar System.

Here we apply \hlll{the} planetary evolution models we developed in \citet{Zuluaga2013}, to study the \gp\ of \Pba, namely planets with masses between 1.3 and $\sim$2.6$\MEarth$, rotational periods between 7.5 and 11.3 days, and habitable surface conditions. Our results are not restricted to \Pb.  The same set of \pa\ we model here, describes well other earth-like exoplanets with similar observable properties (stellar and planetary minimum mass).

\hl{This paper is organized as follows: in Section 2 we describe the planetary evolution model. Section 3 describes the methods we use to estimate the posterior distribution of the \gp\ for \Pba.  Section 4 presents the results of applying our methods and finally, in Section 5 we discuss our results and the limitations of our models.}


\vspace{-0.5cm}
\section{Planetary Evolution Model}
\label{sec:Model}

The planetary evolution model we use here is based in \hl{the model} developed and tested in \citet{Zuluaga2013} (hereafter ZUL13).  \hl{For the sake of completeness and reproducibility, we present in the following paragraphs the main features of the model. We refer the reader to the original paper (especially Section 7) for a detailed discussion on its assumptions, limitations and in particular on the sensitivity of the model to several key parameters.}

\subsection{Interior structure}
\label{subsec:InteriorStructure}

\begin{figure}
  \centering
   \includegraphics[width=0.4
  \textwidth]{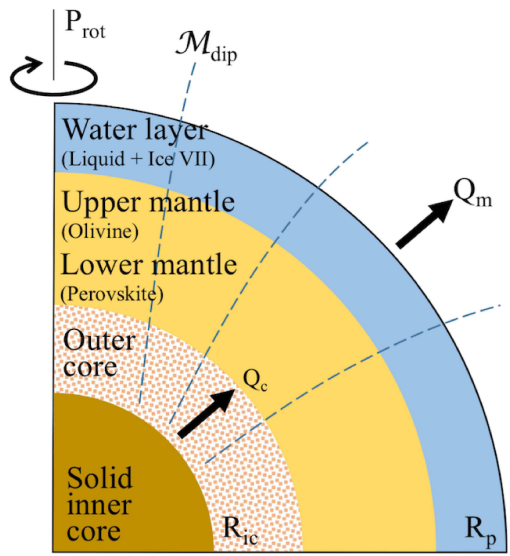}
  \scriptsize
  \caption{Schematic representation of our planetary model. $\Mdip$ is the magnetic dipole moment. $Q\sub{c}$ the heat coming out from the core and $Q\sub{m}$ the total heat released by the planet.\vspace{0.2cm}}
\label{fig:PlanetSchematic}
\end{figure}

Our \Pba\ are solid spherical planets, with three differentiated layer (see Figure \ref{fig:PlanetSchematic} for an schematic representation):
an iron core, a rocky mantle and a surface layer of \hl{solid and liquid} water.  Our original model \hlll{in ZUL13} did not include the \hl{water layer}.  However, in-situ formation of \Pb\ analogues seems to be implausible \citep{AngladaEscude2016} and several alternative formation scenarios \citep{Ribas2016,Barnes2016} predict that the new planet could have a volatile-rich composition \citep{Leger2004}.

Pressure, density, gravitational field and local bulk modulus inside the planet, are calculated solving the hydrostatic equilibrium, continuity and Adams-Williamson equations (\citealt{Valencia06,Valencia07a,Valencia07b}).  For all layers we use the Vinet equation-of-state with parameters frequently used in literature (see Table 1 in ZUL13 \hlll{and \autoref{tab:ThermalModelParameters} here}).  We assume \iac\ that the inner-core is made of pure iron while the outer core is made of an iron alloy, namely Fe$\sub{(0.8)}$FeS$\sub{(0.2)}$. The rocky mantle is made of olivine (upper mantle) and perovskite with ferromagnesiowustite (lower mantle).  \hl{The outermost layer (if present) is made of liquid water and ice (mostly \hlll{ice VII})}.

\begin{figure}
  \centering
   \includegraphics[width=0.5
  \textwidth]{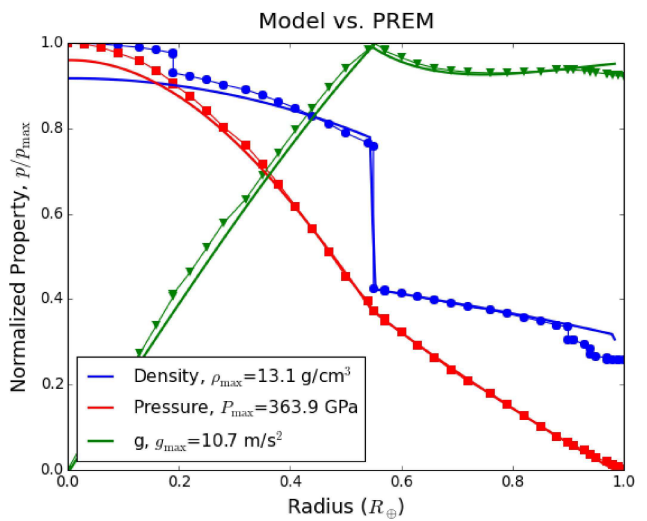}
  \scriptsize
  \caption{\hl{Density, pressure and gravity profiles of an Earth-mass planet as calculated with our interior structure model (continuous lines) and those of the Preliminary Reference Earth Model (PREM, \citealt{Dziewonski1981}).}}
\label{fig:ModelPREM}
\end{figure}

\hl{In Figure \ref{fig:ModelPREM} we show a comparison between selected properties (density, pressure and gravity field) calculated with our model for the case of an Earth-mass planet and the same properties as obtained from seismic data for the Earth itself (the so-called {\it Preliminary Reference Earth Model}, PREM, \citealt{Dziewonski1981}).

Our interior structure model reproduce rather well the mechanical and elastic properties of the Earth's interior. Although the model underestimates the total radius of the planet and the radius of the inner-core, the errors in both quantities are not larger than 2$\%$.}

\hl{The solid inner-core is responsible for the over-density below 0.2 $R_\oplus$ in the PREM data in Figure \ref{fig:ModelPREM}. In our model, on the other hand, iron solidification in the core occurs as the planet evolves. In this case, the density of the solid inner-core (when present) is calculated using the simple prescription $\rho_s(r)=\rho(r)/(1-\Delta \rho)$. Here, $\rho(r)$ is the density obtained solving the interior structure model and $\Delta \rho$ is a constant enhancement factor computed by comparing $\rho(r=0)$ \hlll{with} the density obtained from the Vinet equation \hlll{as calculated with} the pressure in the planet center. In other words, the density profiles shown in Figure \ref{fig:ModelPREM} and \ref{fig:InteriorStructure}, correspond to initial (reference) profiles.}

\hl{

In order to illustrate the effect that different masses and compositions have in planetary properties, we plot in Figure \ref{fig:InteriorStructure} \hll{the} density and pressure profiles of \hlll{several \Pba}.

\begin{figure*}
  \centering

  \vspace{0.2cm}
   \includegraphics[width=70mm]{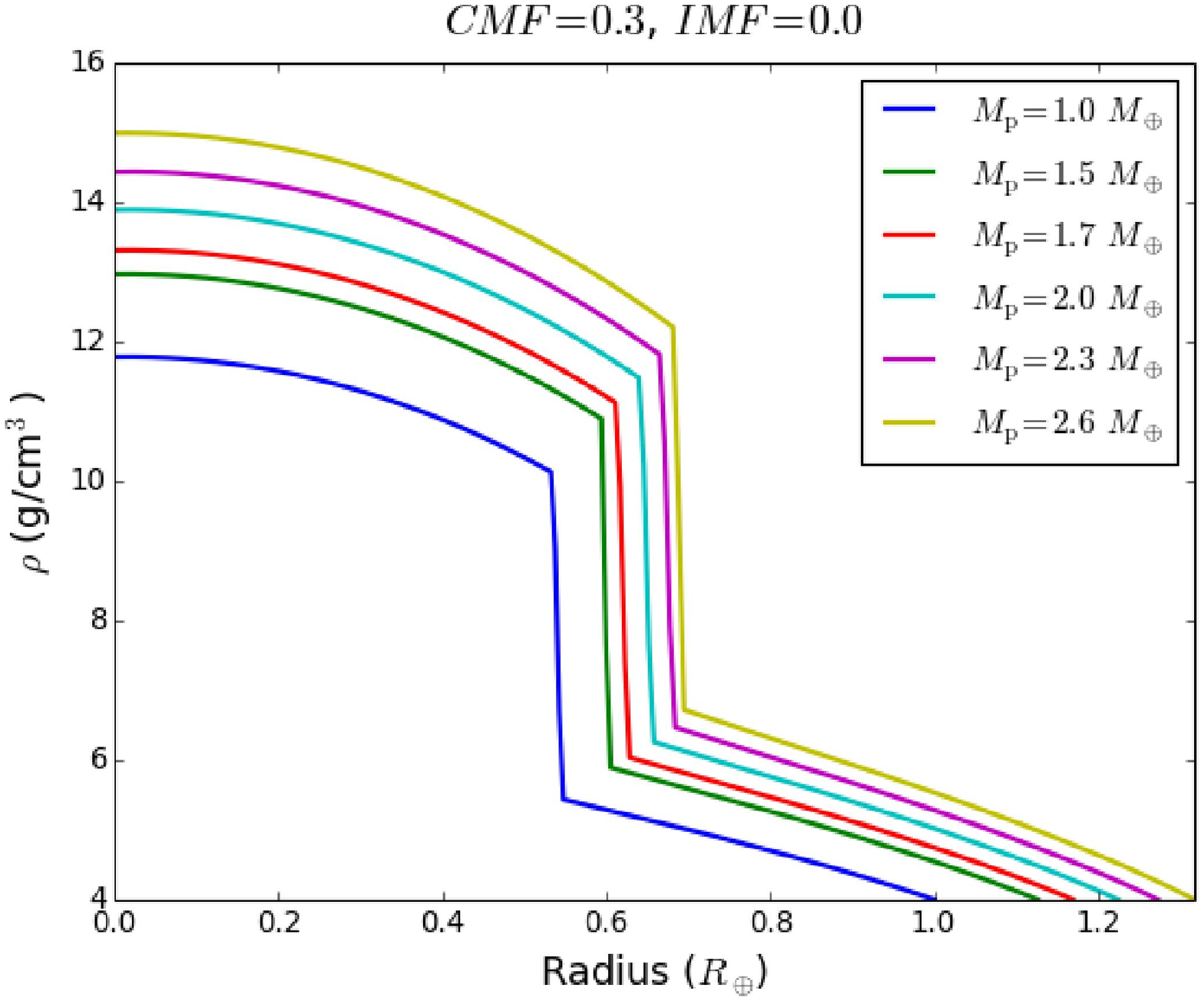}\hspace{0.5cm}
   \includegraphics[width=70mm]{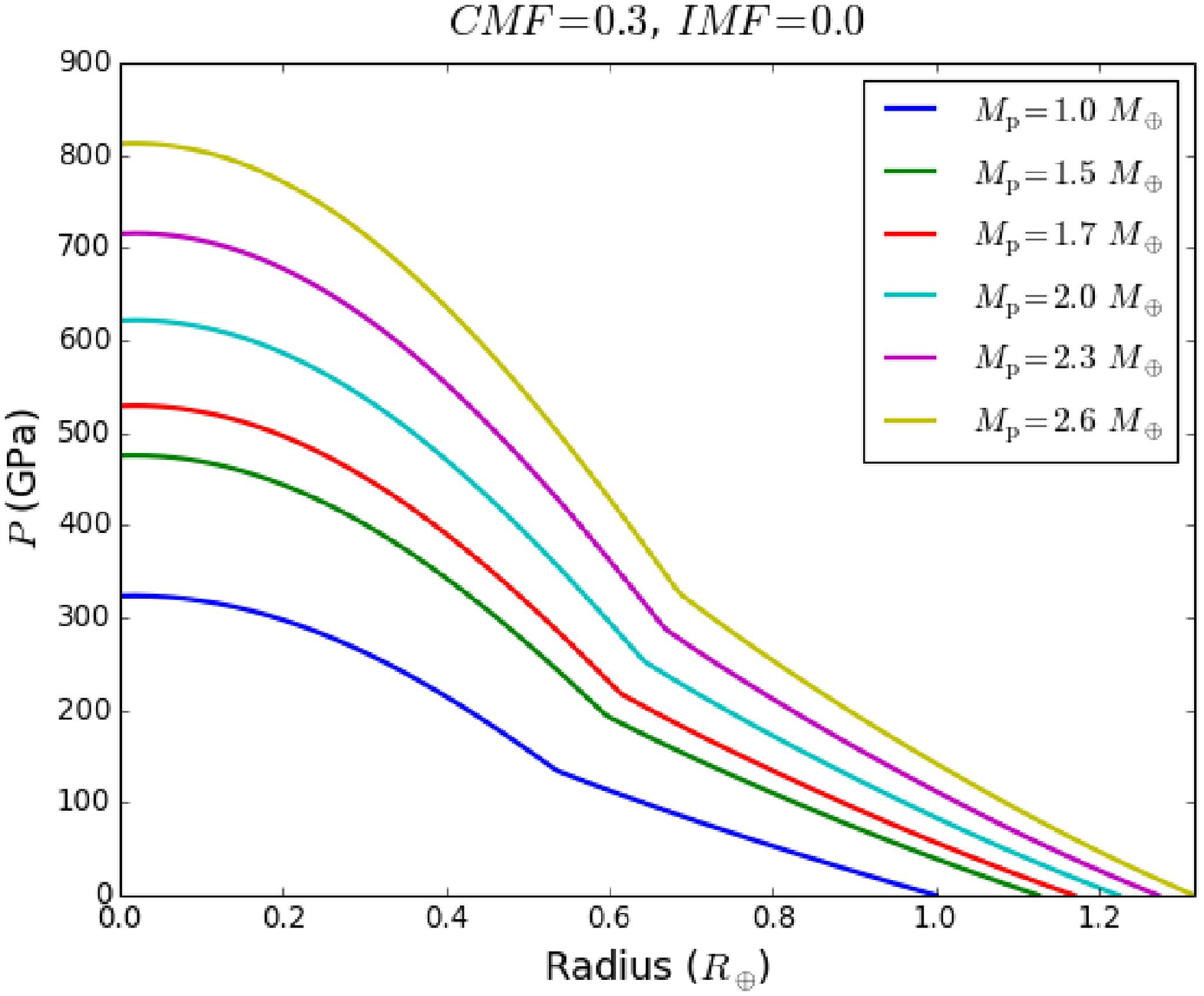}\hspace{0.5cm}\\\vspace{0.3cm}

   \includegraphics[width=70mm]{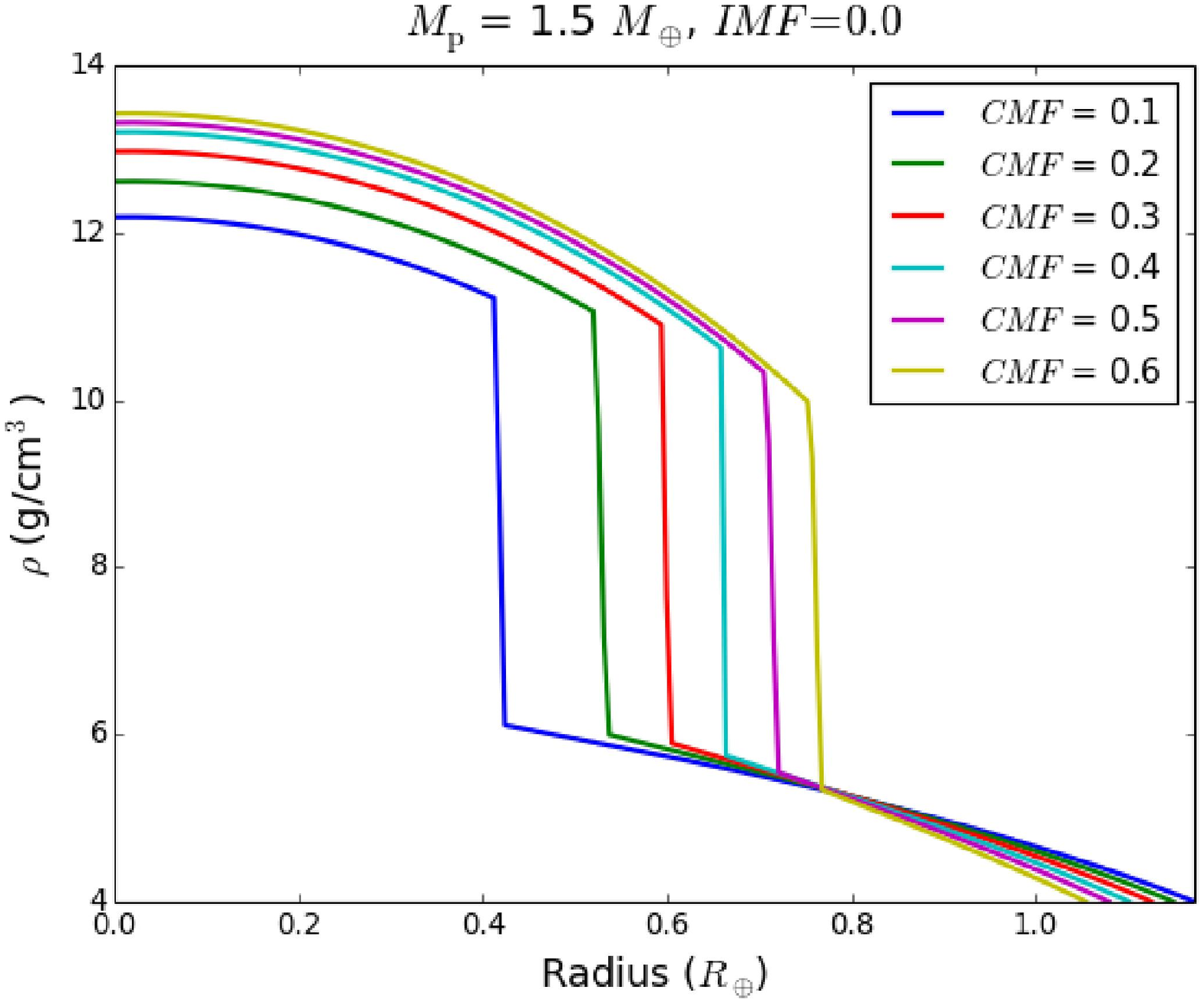}\hspace{0.5cm}
   \includegraphics[width=70mm]{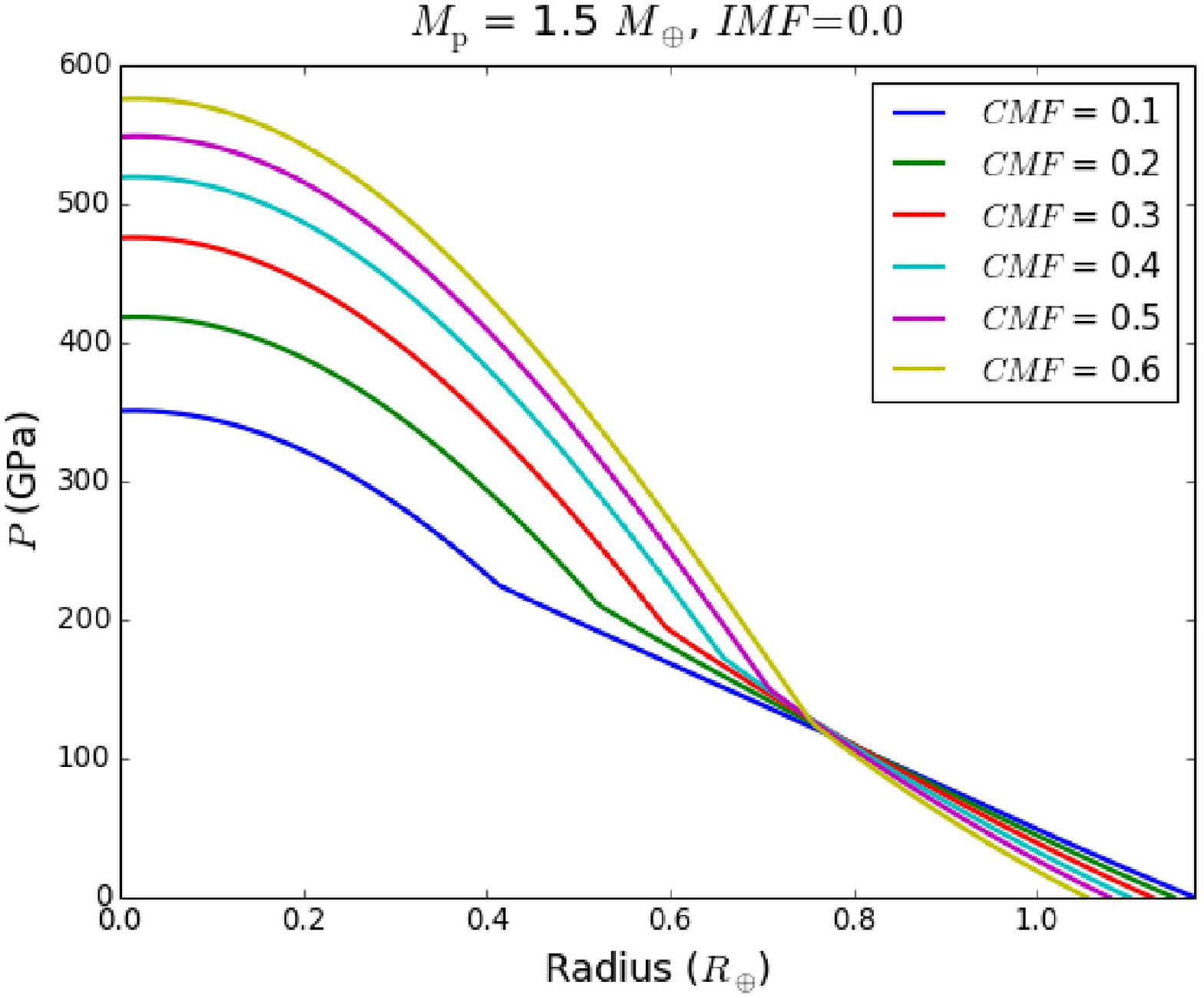}\hspace{0.5cm}\\\vspace{0.3cm}

   \includegraphics[width=70mm]{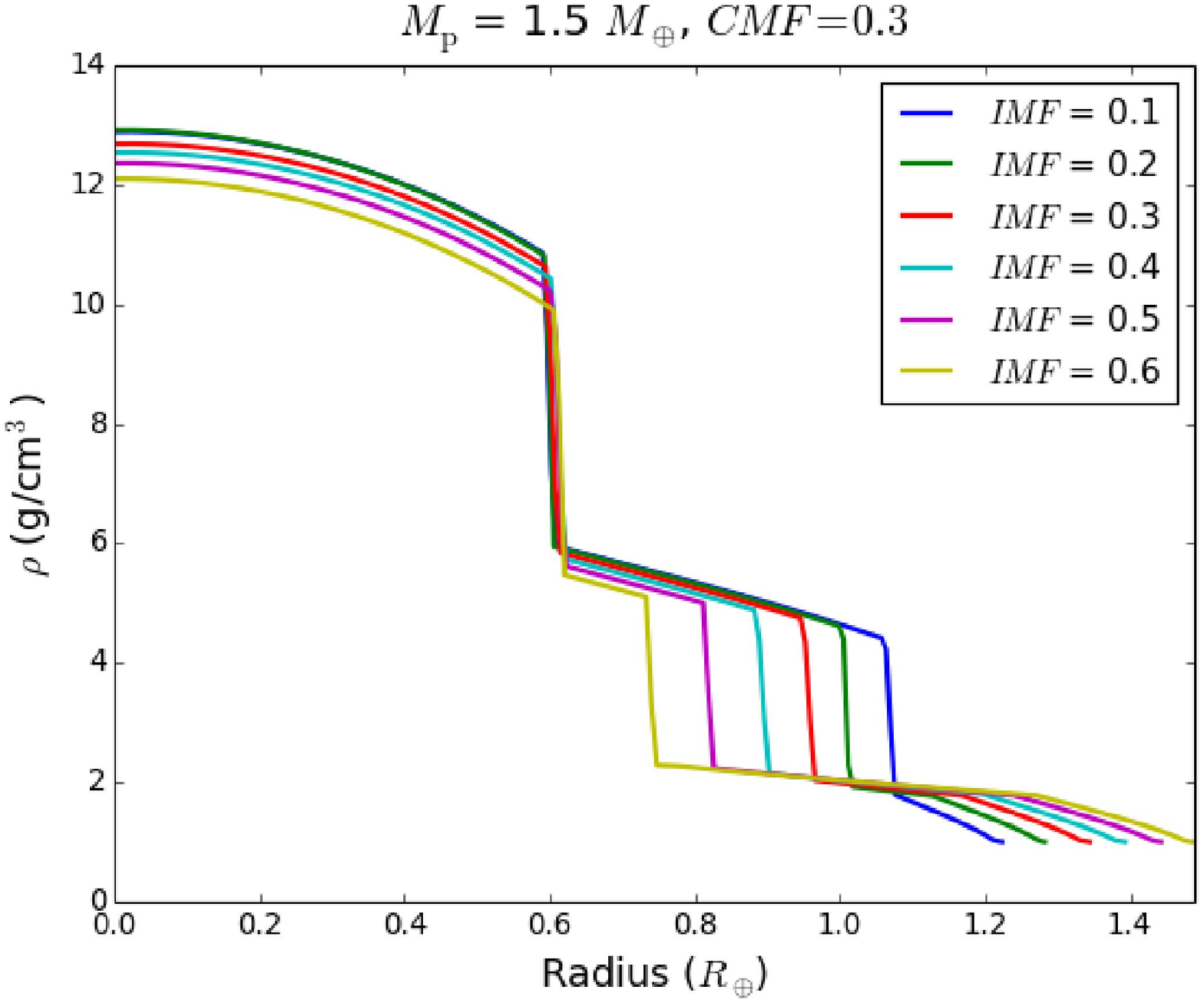}\hspace{0.5cm}
   \includegraphics[width=70mm]{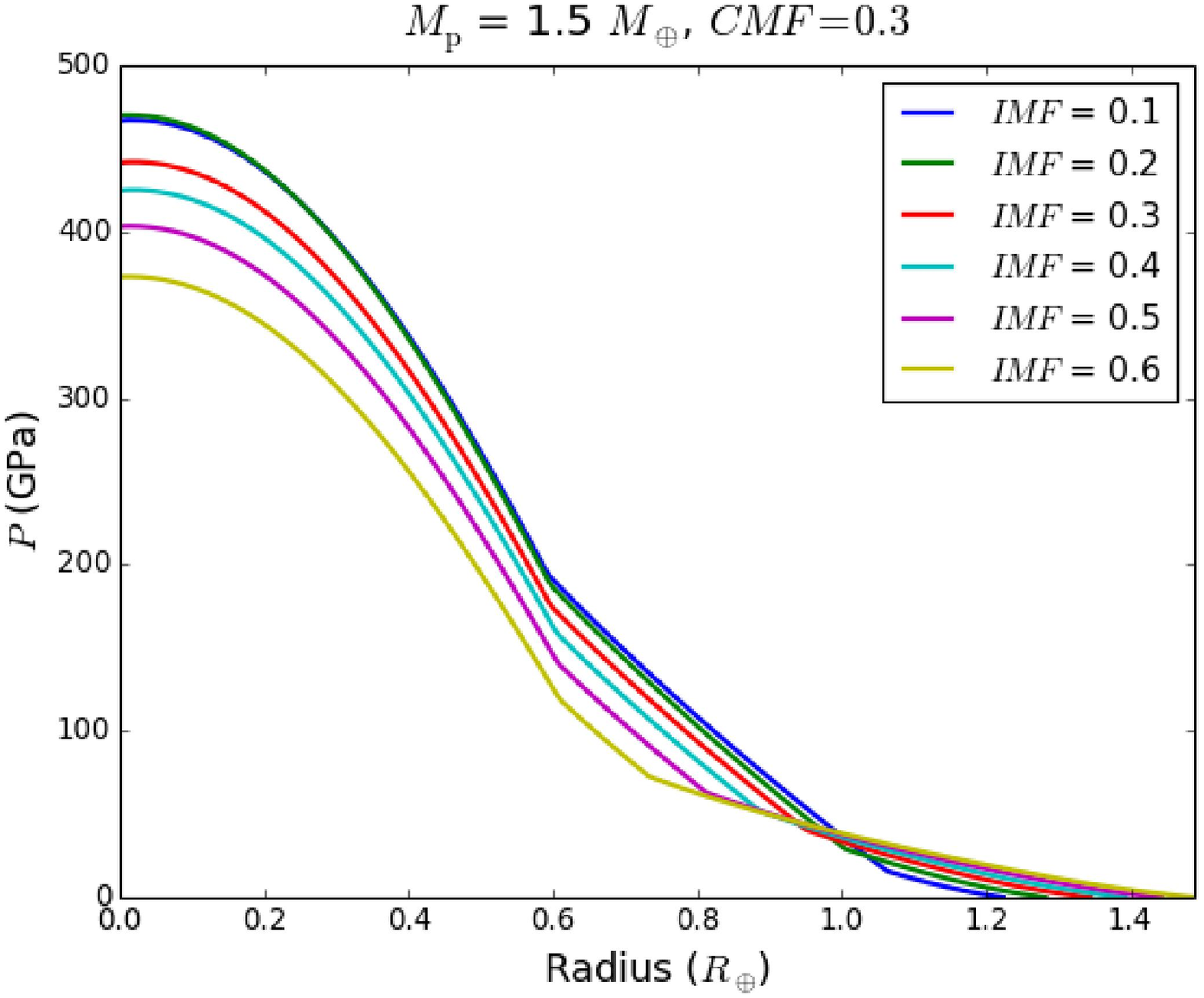}\hspace{0.5cm}\\\vspace{0.0cm}

  \scriptsize
  \caption{\hl{Density and pressure profiles for selected planetary models.  Top row: Earth-like composition and different planetary masses.  Here CMF stands for ``Core Mass Fraction'' and IMF is the ``Ice Mass Fraction''.  Middle row: dry planets ($IMF=0$) with fixed mass $M\sub{p}=1.5\,M_\oplus$ and different iron content. Bottom row: planets with fixed mass $M\sub{p}=1.5\,M_\oplus$, Earth-like iron content $CMF=0.3$ and different amounts of water}. \hlll{The particular models selected here are chosen not to illustrate exhaustively the effect that $M\sub{p}$, CMF and IMF have in the structure of the planet, but to show the effect that each of these quantities could have on the interior properties of \Pba}.\vspace{0.0cm}}
\label{fig:InteriorStructure}
\end{figure*}

As expected, the planetary radius is a monotonous function of planetary mass (upper row of Figure \ref{fig:InteriorStructure}).  For a fixed mass, the radius of the core strongly depends on core mass fraction (CMF) but the total radius of the planet is almost unchanged (middle row of Figure \ref{fig:InteriorStructure}).

In planets having more than 10$\%$ by mass of water (Ice mass fraction, IMF>0.1), the slope of the density profile exhibits several discontinuities.  Those discontinuities arises from phase transitions, namely, liquid to ice and ice to ice.  In Table \ref{tab:WaterParameters} we show the parameters of the equation of state for water we use in our model. Since the pressure in the water layer, at least for the planetary masses we are interested here, is above 1 GPa (see bottom row in Figure \ref{fig:InteriorStructure}) we assume that most ice is only in the form of ice VII (see eg. \citep{Wolanin1997}).

\begin{table}
  \centering
  \scriptsize
  \begin{tabular}{lcccl}
  \hline Water phase & $\rho_0$ (kg/m$^3$) & $K_0$ (GPa) & $K_0'$ \\\hline\hline
  Liquid*  & 998.23 & 2.18 & 0  \\
  Ice VII (1) & 1463 & 27.8 & 2.8 \\
  Ice VII (2) & 1463 & 97.0 & 3.0 \\
  Ice VII (3) & 1463 & 260. & 7.3 \\\hline
  \end{tabular}
\caption{\hl{Parameters of the Vinet Equation of State for water as used in this work \citep{Wolanin1997}.  For liquid water, \hlll{the value} $K_0'=0$ indicates that we assume a constant (density independent) bulk modulus $K_S=K_0$.}\label{tab:WaterParameters}}
\end{table}

We see that in water-rich planets the core and mantle density is not substantially modified with respect to what we can call ``dry planets'' (IMF$\approx$ 0).  Core pressure in water-rich planets, however, is significantly reduced with respect to dry analogues.  This is due to the lower weight of the water-rich external layer.
}

\subsection{Thermal evolution}
\label{subsec:ThermalEvolution}

\hl{A solution to the equations governing the mechanical structure of the planet does not provide its initial temperature profile. Actually, temperature is one of the most uncertain properties when modeling the evolution of solid planets.}

\hl{We will assume that temperature inside the core and mantle follows a an adiabatic profile \citep{Labrosse01,Labrosse03}:}

\hl{
\begin{equation}
\label{eq:AdiabaticProfile}
T(r,t)=T(r=R,t)\exp\left(\frac{R^2 - r^2}{D^2}\right)
\end{equation}
}

\hl{Here $R$ is the outer radius of the corresponding layer (core or mantle radius) and $D$ is the temperature scale-height. In the core $D_c=\sqrt{3 c_p/2\pi\alpha\rho_c G}$, where $\alpha$ is the isothermal expansivity and $\rho_c$ is the density at core center.  In the the mantle $D_m^2=L_m^2/\gamma$, where $L_m^2=(R_p^2-R_c^2)/\log(\rho_m/\rho\sub{CMB})$ is the density scale-height ($\rho_m$ is the mantle average density and $\rho\sub{CMB}$ is the density at the core mantle boundary), and $\gamma$ is the Gruneisen parameter \citep{Labrosse03}.}

\hl{The adiabatic profile in Eq. (\ref{eq:AdiabaticProfile}) explicitly decouple the temporal and spatial dependencies of temperature and reduces the thermal evolution problem to finding two unknowns functions: the upper-mantle temperature $T_m(t)=T(r=R_m,t)$ and the temperature at the core surface $T_c(t)=T(r=R_c,t)$ ($R_c$ and $R_m$ are the mantle and core radii).}

\hl{The initial value of $T_m$ is calculated from the potential temperature $\theta$ of mantle material (Eq. 24 in ZUL13). On the other hand, the initial value of $T_c$ is rather uncertain.  Thermal evolution models available in literature use very different and mostly ``arbitrary'' criteria to set $T_c$ \citep{Tachinami11,Stamenkovic11,Gaidos10,Driscoll2015}.  In ZUL13 we verified that assuming $T_c$ close to the melting point of perovskite at the conditions of the lower mantle, works relatively well at reproducing the thermal properties of an Earth-mass planet (for details see Section 3.4 in ZUL13).  \hll{In this case the initial value is $T_c=4690$ K which is almost 1,000 K larger than the CMB temperature estimated by \citet{Stacey2004}.  We have verified that assuming a larger value for $T_c$ will delay thermal evolution by just 50-100 Myrs which is in agreement with \citep{Stamenkovic11}.}}

\hl{It is interesting to notice that thermal corrections to the equation-of-state should have an impact in the calculation of the mechanical properties of the planet (see eg. \citealt{Valencia07b}).  Strictly speaking we cannot solve the interior structure equations without solving simultaneously the equations governing $T(r,t)$. For simplicity, however, we will neglect thermal corrections to the equations of state.  As a result density, pressure, gravity and other interior structure properties will remain the same as they are in the initial (reference) model during the integration of the thermal evolution equations.}

\hl{Energy conservation across the core-mantle boundary (CMB) and mantle-ice-boundary (MIB) provide us with differential equations governing the evolution of $T_c(t)$ and $T_m(t)$ \citep{Labrosse01,Nimmo09a}:

\begin{eqnarray}
\label{eq:TcEquation}
( C_s + C_g + C_l ) \dtot{T_c}{t} & = & Q_c \\
\label{eq:TmEquation}
Q_r + Q_c + C_m \dtot{T_m}{t} & = & Q_m
\end{eqnarray}
}

\hl{Here $Q_c$ and $Q_m$ are the total energy released by the core and the mantle respectively. $Q_r$ is the radioactive heat produced inside the mantle. $C_s$, $C_g$ and $C_l$ are ``bulk heat capacities'' associated to sensible heat, gravitational energy and latent heat of solidification in the core. Finally $C_m$ is the mantle bulk heat capacity.

The dependency of these quantities on planetary interior properties is rather complex \citep{Nimmo09a}. We summarize below the most relevant details of the calculation of each of these terms. Further information is provided in ZUL13.}

\hl{
{\bf Core $C_s$ and mantle $C_m$ bulk heat capacities}. As the core and the mantle cool down, internal energy is released as heat into the surrounding layers. The amount of energy released by a complete layer after a reduction of one degree Kelvin in their boundary temperature, is given by what we call the ``bulk heat capacity'':

\begin{equation}
C=-4\pi M\int_{R_i}^{R_e} \rho(r)c_p\exp\left(\frac{R^2 -
  r^2}{D^2}\right) r^2 dr
\end{equation}
}

\hl{The minus sign indicates that energy is released to the environment when the temperature of the layer is reduced, ie. $dT<0$.  Here $M$, $R_i$ and $R_e$ are the total mass, inner radius and outer radius of the layer.  $c_p$ is the specific heat capacity of the material and $D$ the temperature scale-height.}

\hl{{\bf Gravitational heat capacity, $C_g$}.  Solidification of iron in the inner\hlll{-}core releases light elements. This material ascends through the outer core bringing out gravitational energy and contributing to cool down the core. To model this effect we use the expression given in Table 1 of \citep{Nimmo09a}.}

\hl{{\bf Latent heat capacity, $C_l$}. Solidification of iron releases latent heat into the core.  To model this effect we use the expression given in Table 1 of \citep{Nimmo09a}.}

\vspace{0.5cm}

\hl{Bulk, gravitational and latent heat capacities requires that the radius of the inner-corem $R_{\rm ic}$ be known at each integration step.  Thus, for instance, if no solid-core is present, ie. $R_{\rm ic}=0$,  $C_g$ and $C_l$ must be set to zero in Eq. (\ref{eq:TcEquation}).

The inner-core radius $R_{\rm ic}$ obeys its own differential equation \citep{Gaidos10}:

$$
\dtot{R_{\rm ic}}{t}=-\frac{D_c^2}{2R_{\rm ic}(\Delta -
  1)}\frac{1}{T_c}\dtot{T_c}{t}.
$$
}

\hl{where $\Delta$ is the ratio between the gradient of the ``solidus''  and the gradient of the adiabatic temperature profile (see ZUL13 for details).}

\hll{For the iron solidus we use the Lindemann law for a Fe+Alloy}:

\beq{eq:IronSolidus}
\der{\log \tau}{\log \rho} = 2[\gamma - \delta(\rho)]
\eeq

\hll{where $\delta(\rho)\approx1/3$ and $\gamma$ is the Gruneisen
parameter that for simplicity is assumed constant. In order to obtain the solidus temperature $\tau$ as a function of $r$, we integrate \autoref{eq:IronSolidus} using reference values
$\rho\sub{0}$ = 8300 kg/m$^3$ (pure iron), $\tau\sub{0} = 1808$ K down to the numerical value of $\rho(r)$ as provided by the interior structure model. Although this solidus depends on the bulk composition of the core (through the Gruneisen parameter), it does not include any composition gradient that may build up during thermal evolution.}

\hl{{\bf Radiogenic heat, $Q_r$}. The decay of unstable isotopes constitutes a significant fraction of the heat released by the interior of solid planets \citep{Kamland2011,Agostini2015}.  Four main isotopes contribute to this effect: K-40, Th-232, U-238 and U-235 \citep{Kite09}.  The total energy released by the decay of those isotopes is given by:

\beq{eq:Qr}
Q_r(t) = M_m \sum_i X_i W_i e^{-\frac{(t-4.5\rm{Gyr})}{t_{1/2}/\log 2}}
\eeq
}

\hl{where $M_m$ is the mantle mass, $X_i$ is the present day concentration of the $i-$th isotope, $W_i$ is the decay specific power and $t_{i,1/2}$  its half-life.  Isotope concentrations depends on the initial composition of the planet and on its impact history.  For the sake of simplicity, we will assume hereafter that all planets have the same radiogenic proportions as Earth \hlll{(see \autoref{sec:Discussion} for a discussion of this particular assumption)}. \hll{The assumed values for the \hlll{parameters in \autoref{eq:Qr}} are presented in \autoref{tab:RadioactiveParameters}}.}

\begin{table}
  \centering
  \begin{tabular}{lcccc}
    \hline Parameter & $^{40}$K & $^{232}$Th & $^{235}$U & $^{238}$U \\\hline\hline
    $t_{1/2}$ (Gyr) & 1.26 & 14.0 & 0.704 & 4.47 \\
    $X_i$ & 36.9 & 124 & 0.22 & 30.8 \\
    $W_i$ ($\times 10^{-5}$ W kg$^{-1}$) & 2.92 & 2.64 & 56.9 & 9.46 \\
    \hline
  \end{tabular}
  \caption{\label{tab:RadioactiveParameters}Parameters of the radioactive decay function assumed in this work.  Taken from Table 1 of \citealt{Kite09} assuming a ``mantle'' composition.}
\end{table}

\vspace{0.5cm}

\hl{Energy balance in Eqs. (\ref{eq:TcEquation}) and (\ref{eq:TmEquation}), depends on the boundary conditions at the core-mantle and mantle-ice boundaries.  We use in both cases boundary layer theory (BLT) to compute the energy flowing through CMB and MBI.}

\hl{{\bf Core and mantle heat, $Q_c$ and $Q_m$}.  According to BLT the heat flowing through a boundary of height $B$ and temperature contrast $\Delta T$ is given by \citep{Gaidos10}:

$$
Q = \frac{4\pi R^2}{B}\,k\,\Delta T\,{\rm Nu},
$$
}

\hl{Here $R$ is the radius of the boundary layer, $k$ is its thermal conductivity and $\Delta T$ the temperature contrast through the boundary.  The Nusselt number, Nu, defined as the ratio of the total to the conductive heat transfer, is expressed in therms of the so-called Rayleigh number Ra \hlll{as}\citep{Gaidos10},}

\hl{
\begin{eqnarray}
\mbox{Nu} & \approx & (\mbox{Ra}/\mbox{Ra}_*)^{\nu\sub{eff}}\\
\mbox{Ra} & = & \frac{\rho\; g\; \alpha\;\Delta T\,B^3}{\kappa \eta}\;\mbox{(h.f.b.)} \\
\mbox{Ra} & = & \frac{\alpha g \rho^2 (Q/M)\,B^5}{k \kappa \eta}\;\mbox{(h.f.i.)}
\end{eqnarray}
}

\hl{\hlll{where Ra$_*$ and $\nu\sub{eff}$ are scaling-law free parameters}, $g$ the local gravitational field, $\kappa$ the thermal diffusivity at the boundary, $Q/M$ the heat per unit of mass coming into the boundary layer, $B$ is the thickness of the layer, and $\eta$ the dynamic viscosity of the material in the boundary layer.} \hlll{Here, {\it h.f.b.} stands for ``heated from below'' and {\it h.f.i.} is ``heated from inside''.}

\hlll{At building our model grid we have used for the parameters of the Nusselt number scaling law, Ra$_*$=1000 and $\nu\sub{eff}=1/3$.  This vale of the exponent is typical of the classical Rayleigh-B\'enard Convection (RPC for short, \citealt{Malkus54a,Malkus54b}).  The proper scaling law for Nu could be much more complex and it would depend on the balance between viscosity, Coriolis force and buoyancy (see eg. \citealt{Chilla2012} and references there in).  In the case of terrestrial planets, rotation may play an important role and taking this factor into account for a proper Nu scaling is important \citep{Gastine2016}.  For the particular case of tidally locked planets (as is the case of Proxima b as we will show later), we have verified that the values of Ra and the convective Rossby number, Ro$\sub{c}$, are close or in the range where the classical non-rotating RBC can be applied.  However, a proper investigation of the effect that a more realistic scaling law for Nu should be pursued.}

\subsubsection{Rhelogical model}

Viscosity is one of the most critical quantities involved in any thermal evolution model.  The election of a particular rheological model to describe the properties of silicates and iron at high pressure has been always a source of controversy (see eg. \citealt{Stamenkovic11}).

After testing different rheological models available in literature, we found that the solution that better matches the thermal evolution of Earth while being also suitable to describe more massive planets (where pressures are much larger than inside the Earth) is to assume two different expressions for computing the dynamic viscosity at low and high pressures.

At high pressures and temperatures (lower mantle and CMB) we use a Nabarro-Herring formula:

\beq{eq:ViscosityModelLower}
\eta_{\rm NH}(P,T) = \frac{R_gd^m}{D_0 A
  m_{mol}}T\rho(P,T)\exp\pr{\frac{b\;T_{\rm melt}(P)}{T}}
\eeq.

\hlll{This model is better suited to describe the rheological properties of perovskite} \citep{Yamazaki01}.
\hll{Here $R_g = 8.31 \mbox{Jmol}^{-1}\mbox{K}^{-1}$ is the gas constant,
$d$ the grain size, $m$ the growing exponent, $D_0$ is the pre-exponential diffusion coefficient, $m_{mol}$ and $T_{\rm melt}(P)$ are the molar density and melting
temperature of Perovskite.  $A$ and $b$ are
free parameters chosen to fit the thermal evolution of Earth.  For all these parameters we have assumed
the same values used in \citet{Stamenkovic11} \hlll{(see \autoref{tab:ThermalModelParameters}).}}

\begin{table*}
  \centering
  \begin{tabular}{llllc}
    Parameter 
    & Value & Ref.\\\hline

    \hline\multicolumn{3}{c}{Bulk}\\\hline
    CMF 
    & 0.325 & -- \\
    $T_s$ 
    & 290 K & -- \\
    $P_s$ 
    & 0 bar & -- \\

    \hline\multicolumn{3}{c}{Inner-core}\\\hline
    Material & 
    & Fe & A \\
    $\rho_0$, $K_0$, $K_0'$, $\gamma_0$, $q$, $\theta_0$ 
    & 8300 kg m$^{-3}$, 160.2 GPa, 5.82, 1.36, 0.91, 998 K & A \\
    $k_c$ 
    & 40 W m$^{-1}$ K$^{-1}$ & B \\
    $\Delta S$ 
    & 118 j kg$^{-1}$K$^{-1}$& C \\

    \hline\multicolumn{3}{c}{Outer core}\\\hline
    Material  
    & Fe$_{(0,8)}$FeS$_{(0,2)}$ & A \\
    $\rho_0$, $K_0$, $K_0'$, $\gamma_0$, $q$, $\theta_0$ 
    & 7171 kg m$^{-3}$, 150.2 GPa, 5.675, 1.36, 0.91, 998 K & A \\
    $\alpha$ 
    & 1.4 $\times 10^{-6}$ K$^{-1}$ & D \\
    $c_p$ 
    & 850 j kg$^{-1}$ K$^{-1}$& C \\
    $k_c$ 
    & 40 W m$^{-1}$ K$^{-1}$ & B \\
    $\kappa_c$ 
    & $6.5\times 10^{-6}$ m$^2$ s$^{-1}$ & E \\
    $\Delta S$ 
    & 118 j kg$^{-1}$ K$^{-1}$& C \\
    $\epsilon_{ad}$ 
    & 0.71 & -- \\
    $\xi_c$  
    & 0.4 & -- \\

    \hline\multicolumn{3}{c}{Lower mantle}\\\hline
    Material 
    & pv+fmw & A \\
    $\rho_0$, $K_0$, $K_0'$, $\gamma_0$, $q$, $\theta_0$ 
    & 4152 kg m$^{-3}$, 223.6 GPa, 4.274, 1.48, 1.4, 1070 K & A \\
    $d$, $m$, $A$, $b$, $D_0$, $m_{mol}$ 
    & 1$\times 10 ^{-3}$ m, 2, 13.3, 12.33, 2.7$\times 10^{-10}$ m$^2$ s$^{-1}$, 0.10039 kg mol$^{-1}$ & F \\
    $\alpha$ 
    & 2.4 $\times 10^{-6}$K$^{-1}$ & D \\
    $c_p$ 
    & 1250 j kg$^{-1}$ K$^{-1}$& C \\
    $k_m$ 
    & 6 W m$^{-1}$ K$^{-1}$ & C \\
    $\kappa_m$ 
    & $7.5\times 10^{-7}$ m$^2$s$^{-1}$ & E \\
    $\Delta S$ 
    & 130 j kg$^{-1}$ K$^{-1}$& C \\

    \hline\multicolumn{3}{c}{Upper mantle}\\\hline

    Material 
    & ol & A \\
    $\rho_0$, $K_0$, $K_0'$, $\gamma_0$, $q$, $\theta_0$ 
    & 3347 kg m$^{-3}$, 126.8 GPa, 4.274, 0.99, 2.1, 809 K & A \\
    $B$, $n$, $E^*$, $\dot \epsilon$ 
    & $3.5\times 10 ^{-15}$ Pa$^{-n}$s$^{-1}$, 3, 430$\times 10^{3}$ j mol$^{-1}$, $1\times 10 ^{-15}$ s$^{-1}$ & D \\
    $V^*$ 
    & 2.5 $\times 10^{-6}$ m$^3$ mol$^{-1}$ & F \\
    $\alpha$ 
    & 3.6 $\times 10^{-6}$ K$^{-1}$ & D \\
    $c_p$ 
    & 1250 j kg$^{-1}$ K$^{-1}$& C \\
    $k_m$ 
    & 6 W m$^{-1}$ K$^{-1}$ & C \\
    $\kappa_m$ 
    & $7.5\times 10^{-7}$ m$^2$ s$^{-1}$ & E  \\
    $\theta$ 
    & 1700 K  & F \\
    $\chi_r$ 
    & 1.253 & -- \\

    \hline
  \end{tabular}
  \caption{Interior structure and thermal evolution model
    parameters (for a \hlll{detailed definition of these quantities} see \citealt{Zuluaga2013}. Sources: (A) \citet{Valencia06}, (B) \citet{Nimmo09a},
    (C) \citet{Gaidos10}, (D) \citet{Tachinami11}, (E)
    \citet{Ricard09}, (F) \citet{Stamenkovic11}.\label{tab:ThermalModelParameters}}
\end{table*}

At low pressures and temperatures (upper mantle), where rheology is relatively better constrained, the Nabarro-Herring formula leads to huge underestimations of viscosity. In this case we use a more usual Arrhenius-type formula \citet{Tachinami11}:

\beq{eq:ViscosityModelUpper}
\eta_A(P,T) = \frac{1}{2}\cor{ \frac{1}{B^{1/n}}\exp\pr{ \frac{E^* +
      PV^*}{nR_gT} } }\dot{\epsilon}^{(1-n)/n}
\eeq

Here $\dot{\epsilon}$ is the strain rate, $n$ is the creep index, $B$
is the Barger coefficient, and $E^*$ and $V^*$ are the activation
energy and volume.  The values assumed here for these parameters are
the same as those given in table 4 of \citealt{Tachinami11} except for
the activation volume whose value is assumed to be $V^*=2.5\times
10^{-6}$ m$^3$ mol$^{-1}$.

Although still controversial, our approach is physically better motivated than other solutions to the same problem \hll{(that for instance use the same rheology all across the mantle)}.  Naturally it could be subject to huge improvements as better rheological models at high pressures be developed in the future.

\begin{figure}
  \centering
   \includegraphics[width=0.5
  \textwidth]{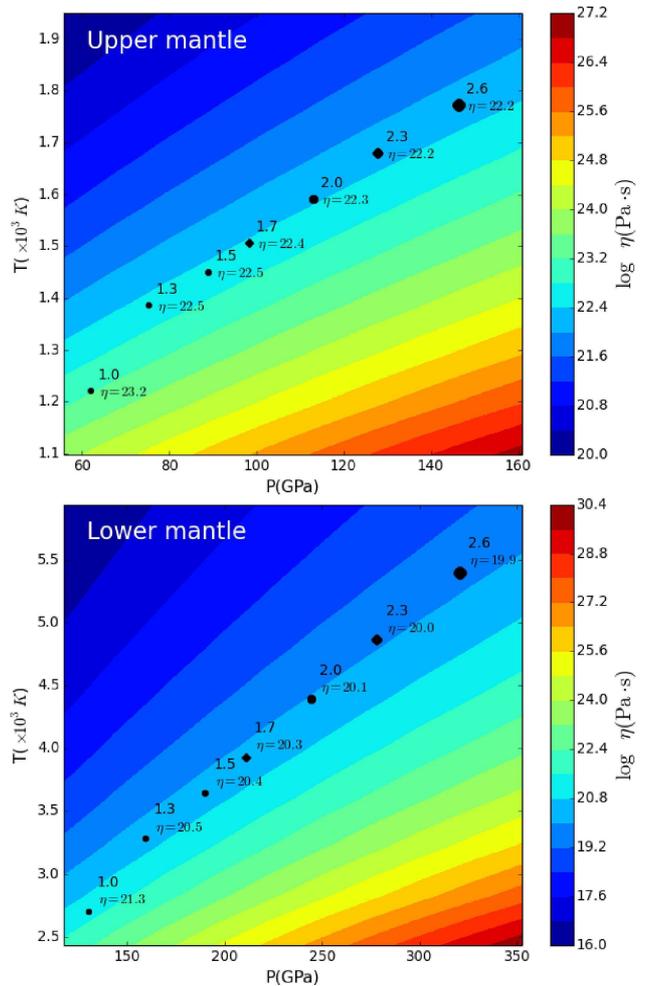}
  \scriptsize
  \caption{\hll{Values of the viscosity as calculated in our model for the lower mantle (Nabarro-Herring model) and the upper-mantle (Arrhenius-like model).  Black circles correspond to values of (upper or lower mantle) pressure, temperature and resulting viscosity in planets with different mass and same composition as Earth, as computed at the middle of the dynamo-lifetime.}}
\label{fig:Viscosity}
\end{figure}

\hll{In \autoref{fig:Viscosity} we plot contour values of the viscosity at the typical pressures and temperatures in the interior of the planetary models in \autoref{fig:InteriorStructure}.  We see that the viscosity values used in our models are in good agreement with those obtained from glacial isostatic adjustment data \citep{Mitrovica2004,Lau2016} and those assumed in previous thermal evolution models \citep{Stamenkovic11}.}

\subsubsection{Thermal evolution results}

\hll{In Figure \ref{fig:ThermalEvolution} we show the heat flux, $q_m=Q_m/4\pi R_p^2$ at the planet surface calculated for the same set of planetary models of Figure \ref{fig:InteriorStructure}.  We also plot there the total available energy for dynamo action $Q_{\rm conv}$}:

\begin{figure*}
  \centering

  \vspace{0.2cm}
   \includegraphics[width=70mm]{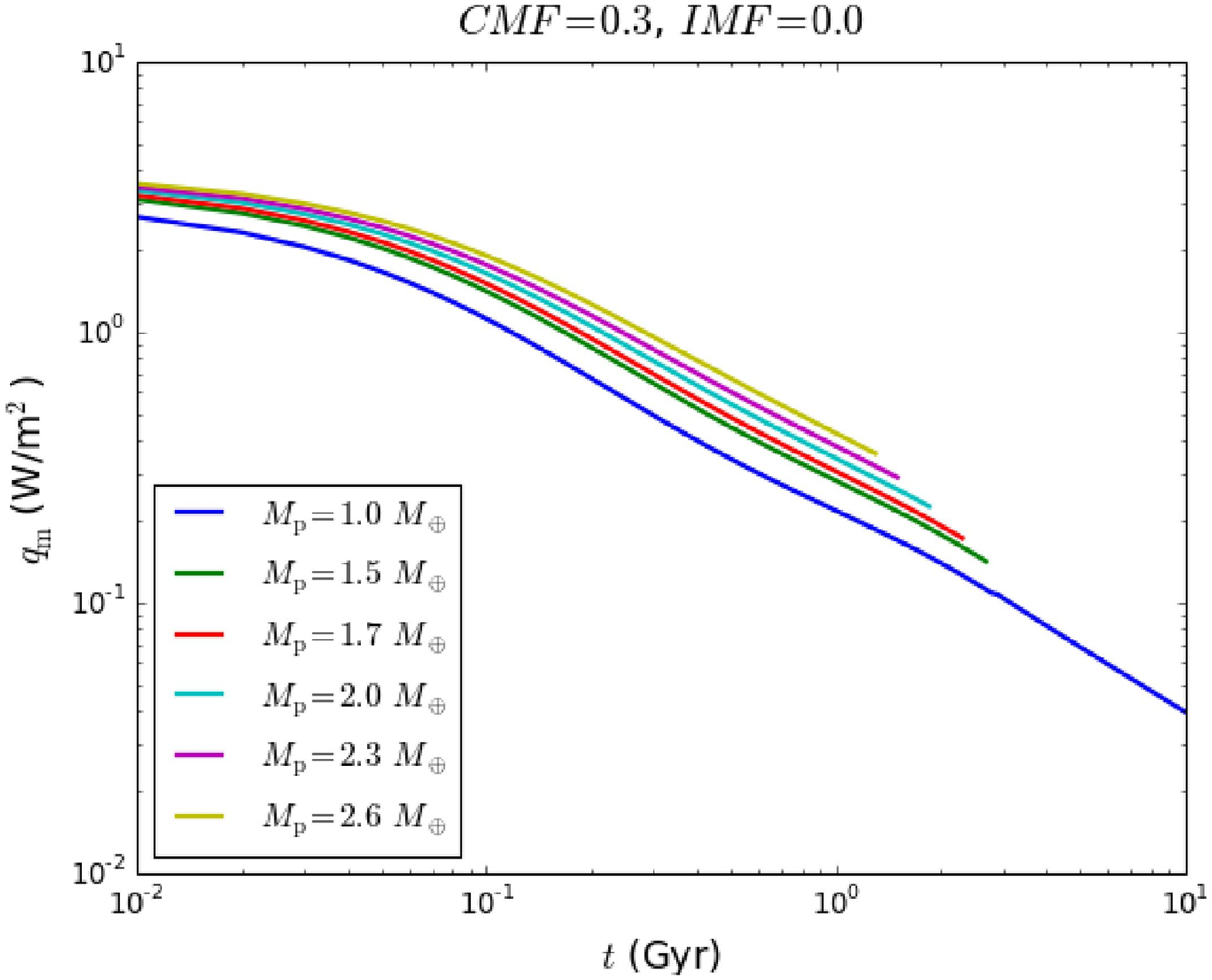}\hspace{0.5cm}
   \includegraphics[width=70mm]{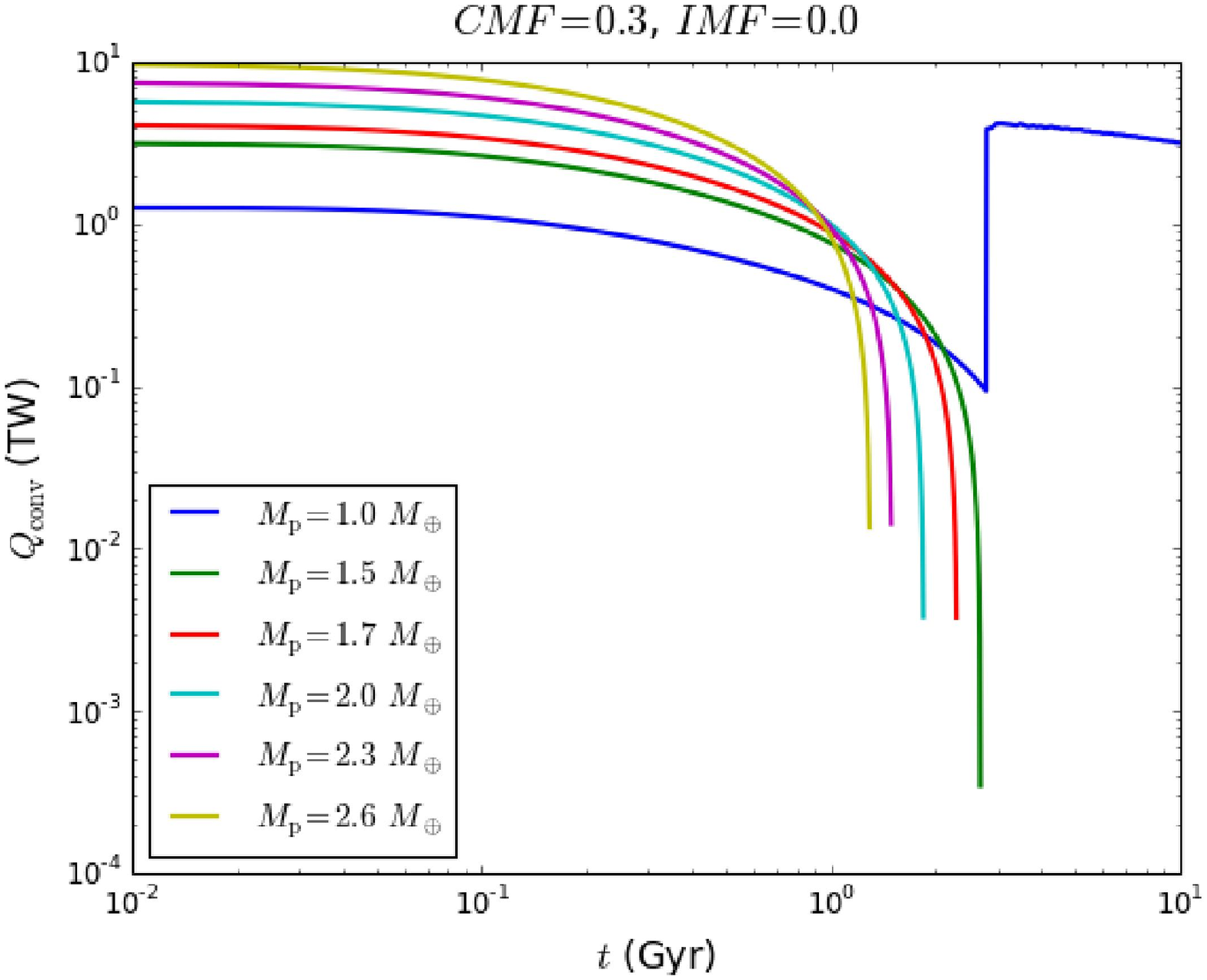}\hspace{0.5cm}\\\vspace{0.3cm}

   \includegraphics[width=70mm]{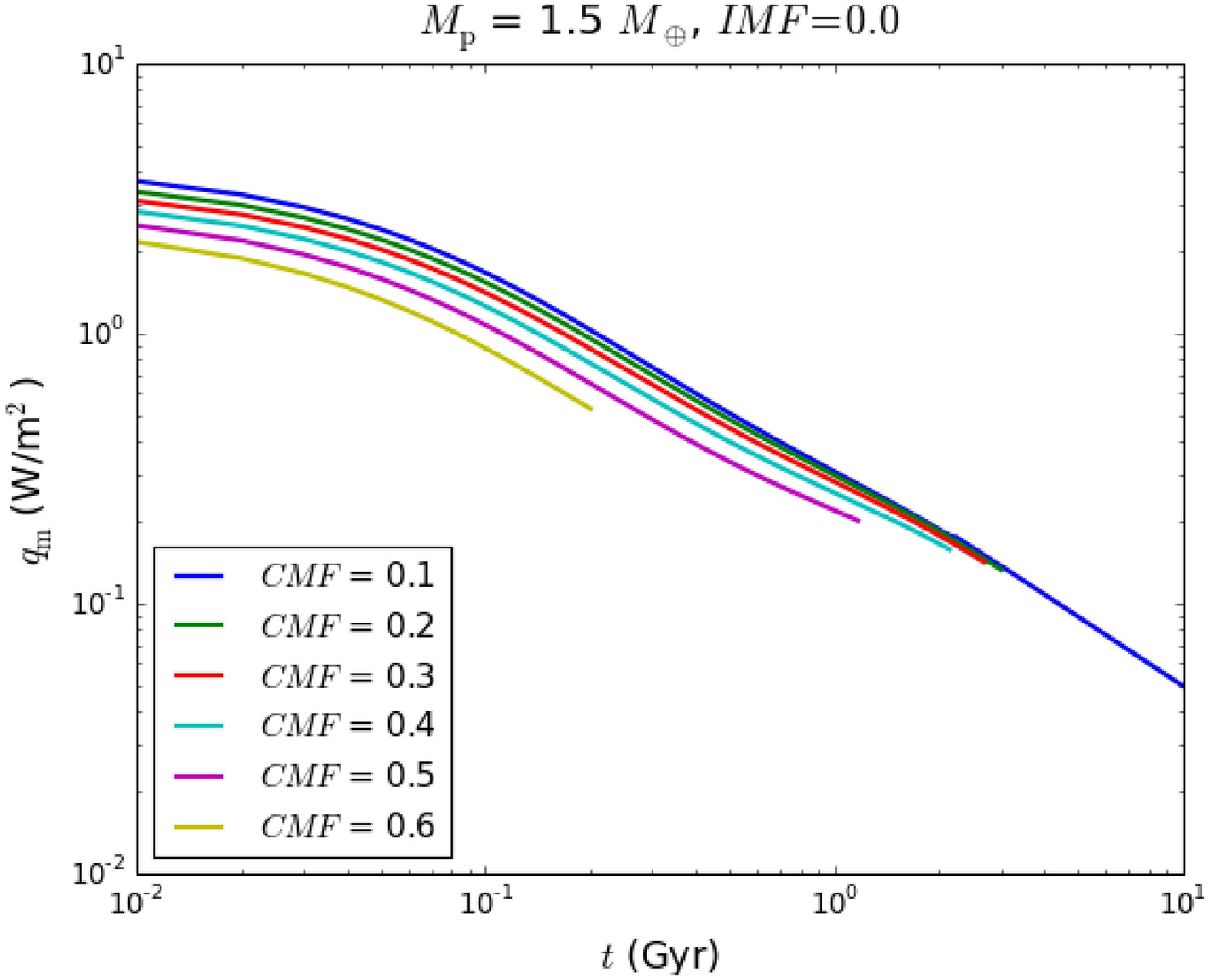}\hspace{0.5cm}
   \includegraphics[width=70mm]{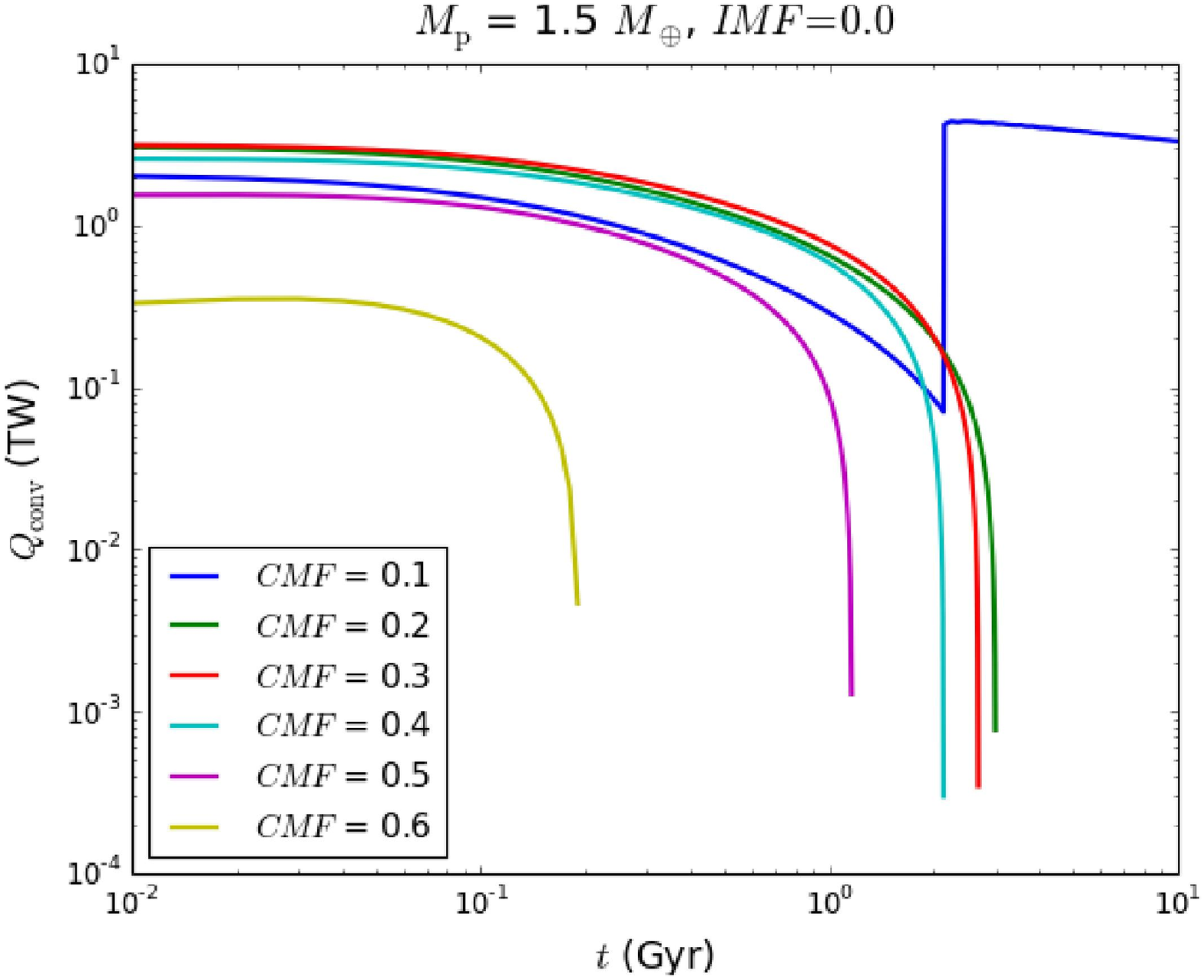}\hspace{0.5cm}\\\vspace{0.3cm}

   \includegraphics[width=70mm]{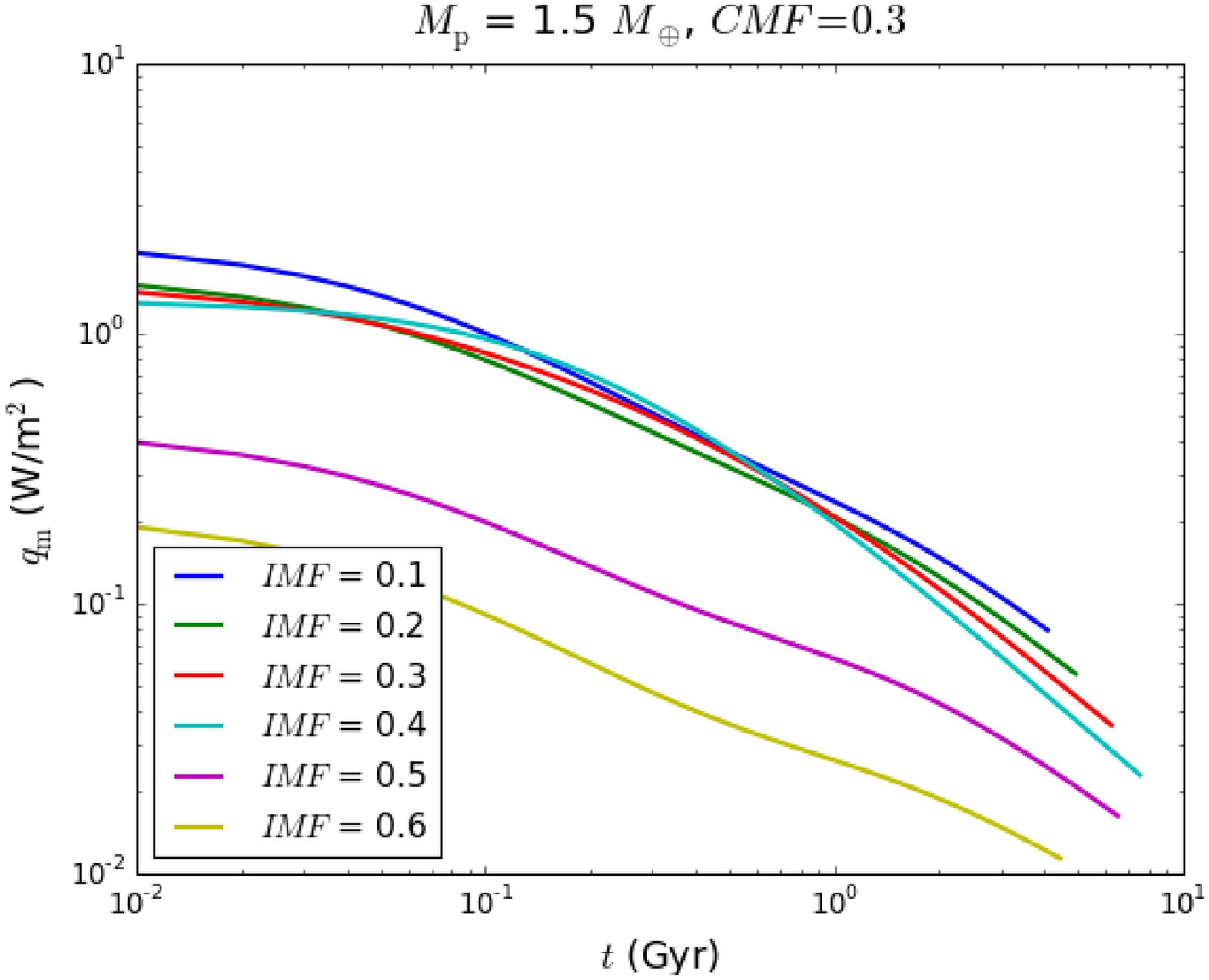}\hspace{0.5cm}
   \includegraphics[width=70mm]{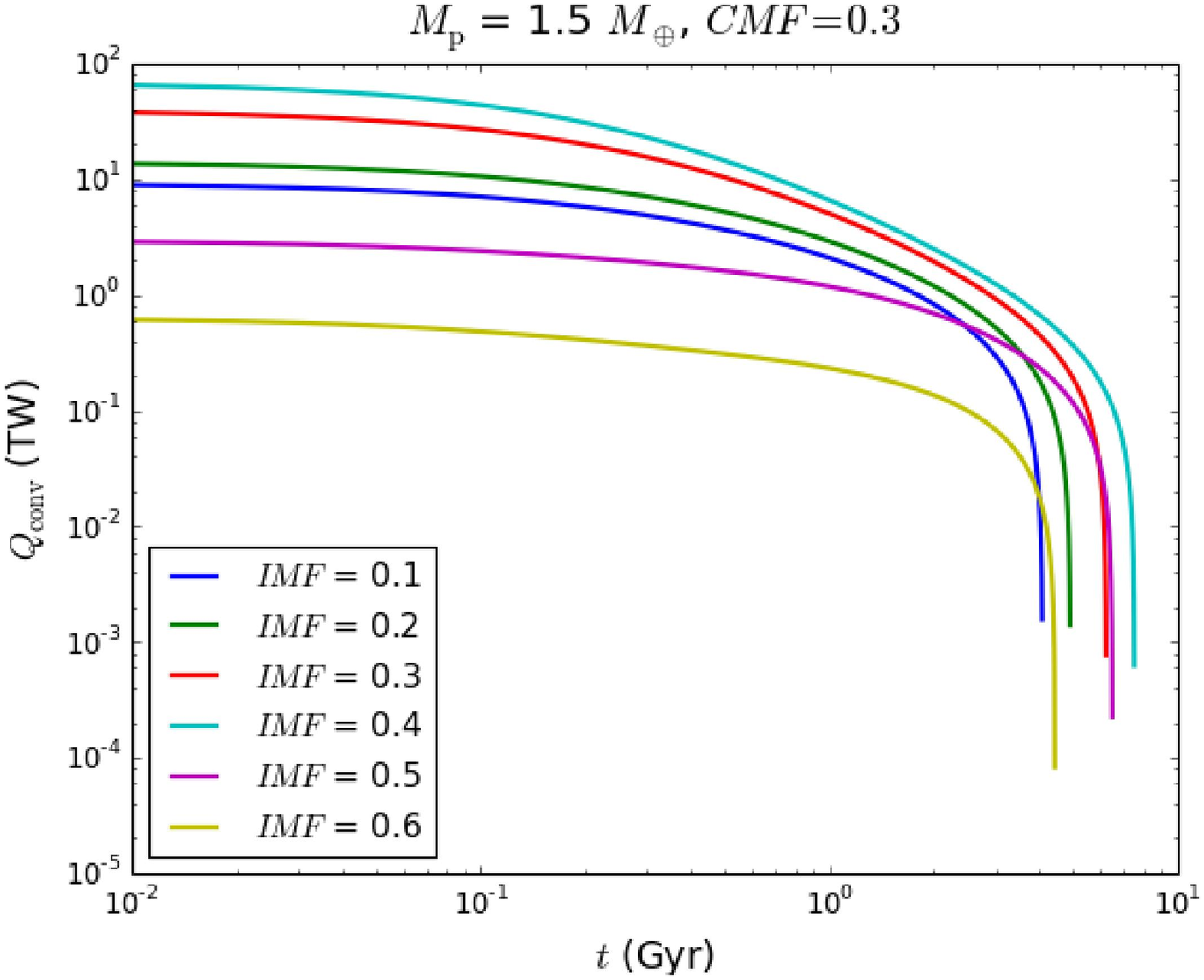}\hspace{0.5cm}\\\vspace{0.0cm}

  \scriptsize
  \caption{\hl{Surface heat flux $q_m$ and total \hll{convection energy} in the core $Q_{\rm conv}$ for the same planetary models used in Figure \ref{fig:InteriorStructure}}. \hlll{Values for each model are plotted until dynamo is shut down, ie. $Q\sub{conv}<0$; we call this time, the dynamo lifetime $t\sub{dyn}$.  Of course the mantle still cools down beyond $t\sub{dyn}$; however, since our model is intended to calculate the magnetic properties of the planet, the simulation is stopped at that time}.\vspace{0.0cm}}
\label{fig:ThermalEvolution}
\end{figure*}

\beq{eq:EntropyBalance}
Q\sub{conv}=\Phi T\sub{c}
\eeq

\hll{where $\Phi$ is the entropy associated to Ohmic dissipation:}

\beq{eq:EntropyBalance}
\Phi = E_l +  E_s + E_g - E_k.
\eeq

\hll{$E_s$, $E_g$, $E_l$ and $E_k$ are the entropies associated to secular cooling, compositional buoyancy, latent heat and thermal conduction (thermal diffusivity as ohmic dissipation is a sink of entropy and hence have the same sign as $\Phi$). We have neglected here the terms from radioactive and pressure heating \citep{Nimmo09a}.}

\hll{Analytical expressions for the sources of entropy, $E_s$, $E_g$ and $E_l$, in the case of a core having adiabatic temperature and density profiles are given in Table 1 of \citep{Nimmo09b}.} \hll{On the other hand, the entropy associated to thermal conduction is given by:}

$$
E_k=\frac{16\pi\kappa r\sub{c}^5}{5D^4}
$$

\hll{where $\kappa$ is the thermal conductivity.  In this work we have assumed an outdated value of $\kappa=40$ W m$^{-1}$ K$^{-1}$, the same used in the original work of ZUL13 to fit the thermal evolution of the Earth. Larger values for this parameter, $\kappa>90$ W m$^{-1}$ K$^{-1}$ have been recently predicted by {\it ab initio} calculations \citep{Pozzo12,Gomi2013}.}  However, \hll{In ZUL13 we studied the effect that a larger $\kappa$ will have on the results of our thermal evolution model.  We found there that doubling $\kappa$ will not have a significant impact on the maximum magnetic field strength or the dynamo lifetime, at least for planets with a similar composition as the Earth.  For more massive planets (and probably for planets having a large water envelop), using a lower value of $\kappa$ tend to overestimate the dynamo lifetime. We must have this in mind when interpreting the constraints on the magnetic properties of Proxima b analogues.  Running the full model grid for larger values of $\kappa$ is left for a future work.}

\hl{Thermal evolution simulations end when \hll{the total heat flux is equal to the conductive flux (no convection)}, and the dynamo is shut down.  The maximum simulation time is called the dynamo lifetime $t_{\rm dyn}$.}

\hl{Surface heat flux is not too different among dry planets (IMF$\approx 0$) having different masses and compositions (upper row in Figure \ref{fig:InteriorStructure}).  However, the presence of a water layer, although not contributing to the energy budget of the planet, have a large impact on the thermal output and core convective power.  Water-rich planets produce up to one order of magnitude less energy than dry ones.  This is due to their thinner and less massive rocky mantles.

Interestingly, the core convective power is larger for larger IMF (at least until IMF$\approx 0.4$ in the case of the modeled mass, see bottom row in Figure \ref{fig:InteriorStructure}). This is due to a thinner and cooler mantle that ensues the removal of heat from the core, boosting convection.  However, if the mass of the water layer is too large, the pressure and temperature in the core is low, as well as its size and mass (see bottom row in Figure \ref{fig:InteriorStructure}) reducing the power of convection.

Water-rich planets also have longer-lived dynamos than drier ones (middle and bottom row in Figure \ref{fig:ThermalEvolution}).  This is due to the fact that the cooling rate $dT/dt$ is proportional to the total heat released by the planet (see Eqs. \ref{eq:TcEquation} and \ref{eq:TmEquation}). Since water rich planets cool down slower, their dynamos could also last longer.
}

\subsection{Magnetic properties}
\label{subsec:MagneticProperties}







Energy-based \hl{dynamo} scaling laws \citep{Christensen10} \hl{allows us to} estimate from the available convective energy, the volumetric energy density of the magnetic field in the core \hl{$B\sub{rms}^2=(1/V)\int B^2 dV$}:

\hl{
\begin{equation}
B\sub{rms}\propto \mu_0^{1/2} {\bar{\rho}_c}^{1/6} (D/V)^{1/3}
Q\sub{conv}^{1/3}
\label{eq:Brms}
\end{equation}
}

\hl{Here $\bar{\rho}_c$, $D=R_\star-R\sub{ic}$ and
$V=4\pi (R_\star^3-R\sub{ic}^3)/3$ are the average density, height
and volume of the convective shell.}

\hl{Magnetic energy} could be distributed in many multipolar components (multipolar-dominated field) or concentrated in the dipole component (dipolar-dominated field).  \hl{\hlll{The fraction of the power in the dipolar component or the ``degree of dipolarity''}, depends on different factors, including convective power, density and size of the convective layer, but more importantly on planetary rotational period $P$.}

\hl{To determine \hlll{the degree of dipolarity}, we need to estimate the so-called Rossby number; this quantity is also scaled from the convective energy using a power-law \hlll{\citep{Aubert09,Zuluaga2013}}}:

\beq{eq:Rol}
Ro^*_l \propto {\bar{\rho}_c}^{-1/6} R_c^{-2/3} D^{-1/3} V^{-1/2}
Q\sub{conv}^{1/2} P^{7/6}.
\eeq

\hl{In numerical experiments dynamos with $Ro^*_l\lesssim 0.1$ have dipolar dominated magnetic fields.}

\hl{Once the average intensity and \hlll{the degree of dipolarity} of the core magnetic field has been estimated, we calculate an upper bound of the core dipolar field $B^{\rm dip}_c$.  We use for that purpose the prescription in \citet{Zuluaga12}, \hlll{for which} we first calculate the maximum fraction of the magnetic energy $1/b\sub{dip}^{min}$ in the dipolar component of the field (see Figure 1 in \citealt{Zuluaga12}), \hlll{and from there to constrain the dipolar component of the core field,}}

\beq{eq:Bdipmax_scaling}
B_{c}^{dip}\lesssim \frac{1}{b\sub{dip}^{min}(Ro^*_l)} B\sub{rms}
\eeq

\hl{Finally, the maximum attainable dipolar magnetic field strength at the planetary surface is computed from:}

$$
B_p^{\rm dip}(R_p)=B_c^{dip}\left(\frac{R_p}{R_c}\right)^3
$$

\hl{The strength of the magnetic field can also be expressed in terms of the dipole moment $\Mdip$, which is related to the dipolar component of the magnetic field $B^{\rm dip}$ by:}

\hl{
$$
\Mdip = \frac{2\sqrt{2}}{\mu_0}\pi R_p^3 B^{\rm dip}_p
$$
}

\hl{where $\mu_0$ is the vacuum permeability.} \hl{For reference, the present value the Earth's surface magnetic field dipole component is $B^{\rm dip}_\oplus=42.48$ $\mu$T, while its dipole moment is $\MdipEarth=7.768\times 10^{22}$ A m$^2$.}

\begin{figure*}
  \centering

  \vspace{0.2cm}
   \includegraphics[width=70mm]{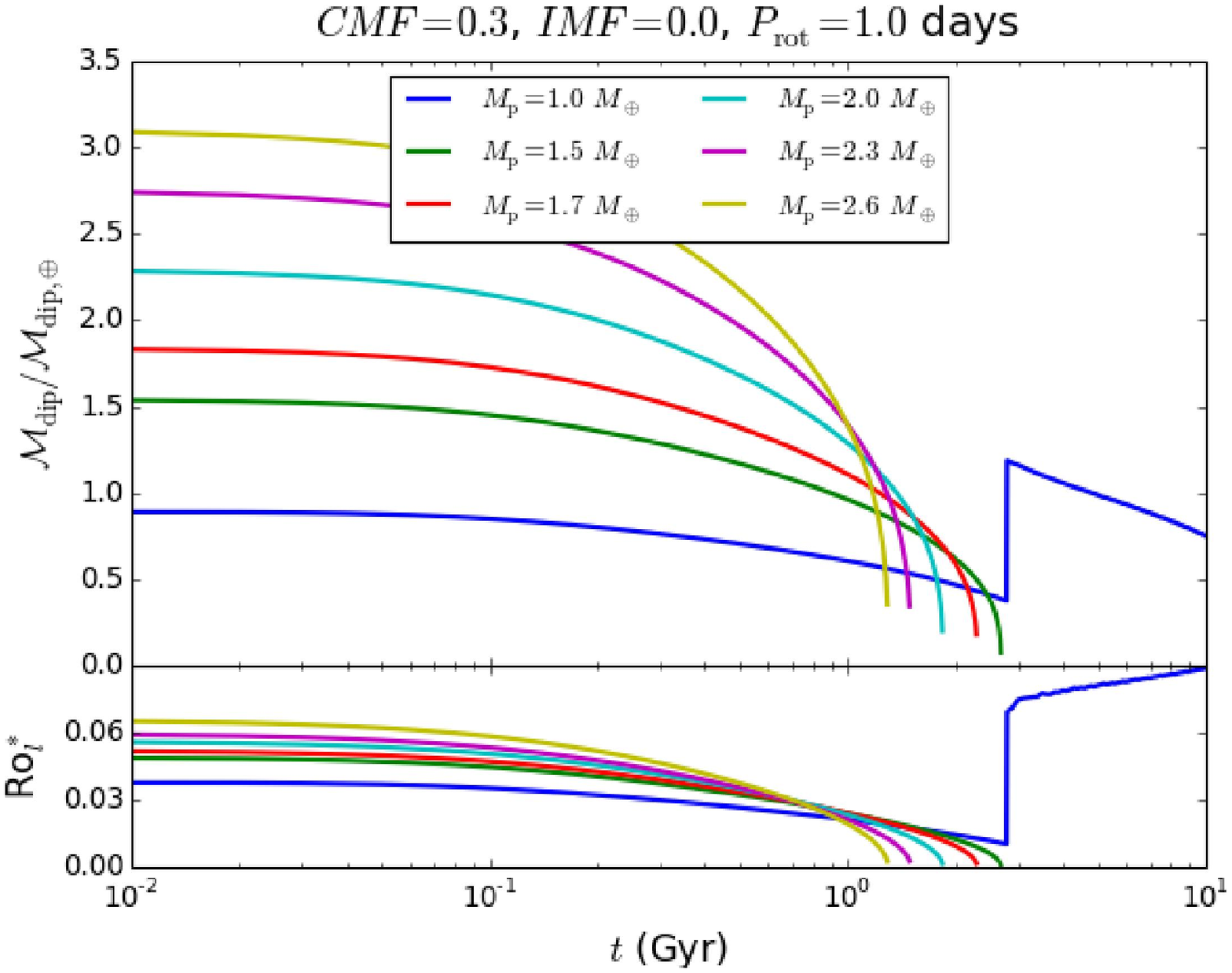}\hspace{0.5cm}
   \includegraphics[width=70mm]{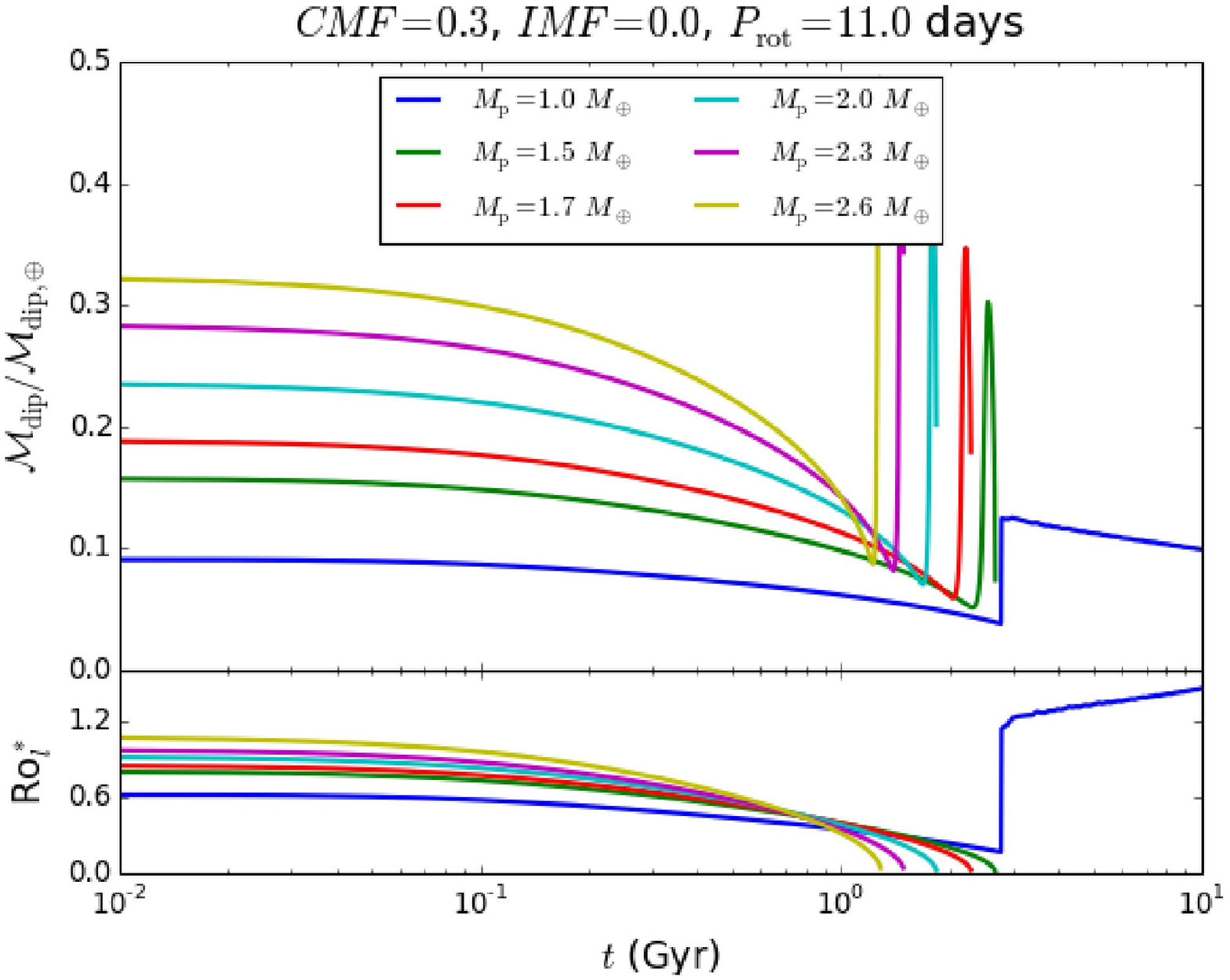}\hspace{0.5cm}\\\vspace{0.3cm}

   \includegraphics[width=70mm]{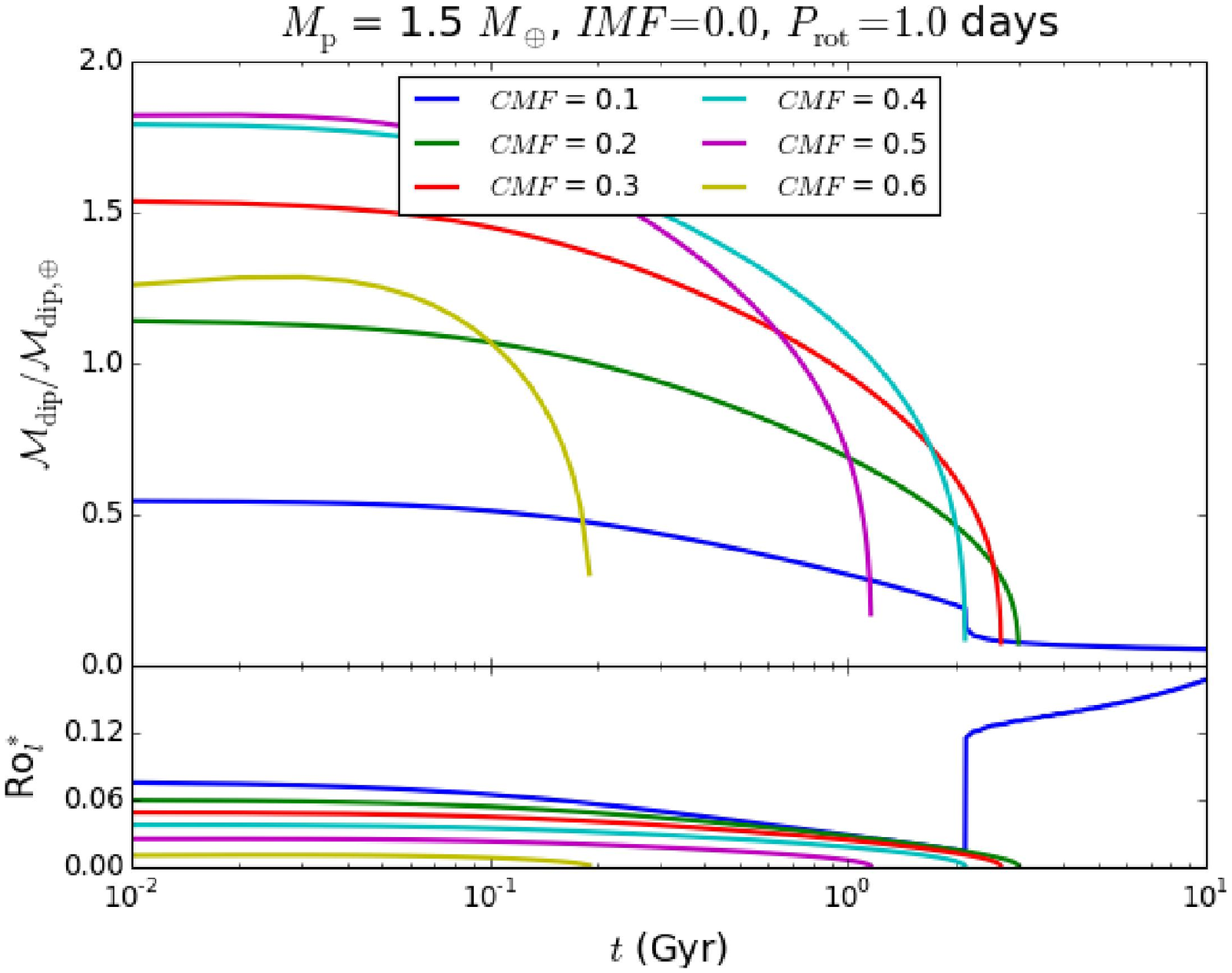}\hspace{0.5cm}
   \includegraphics[width=70mm]{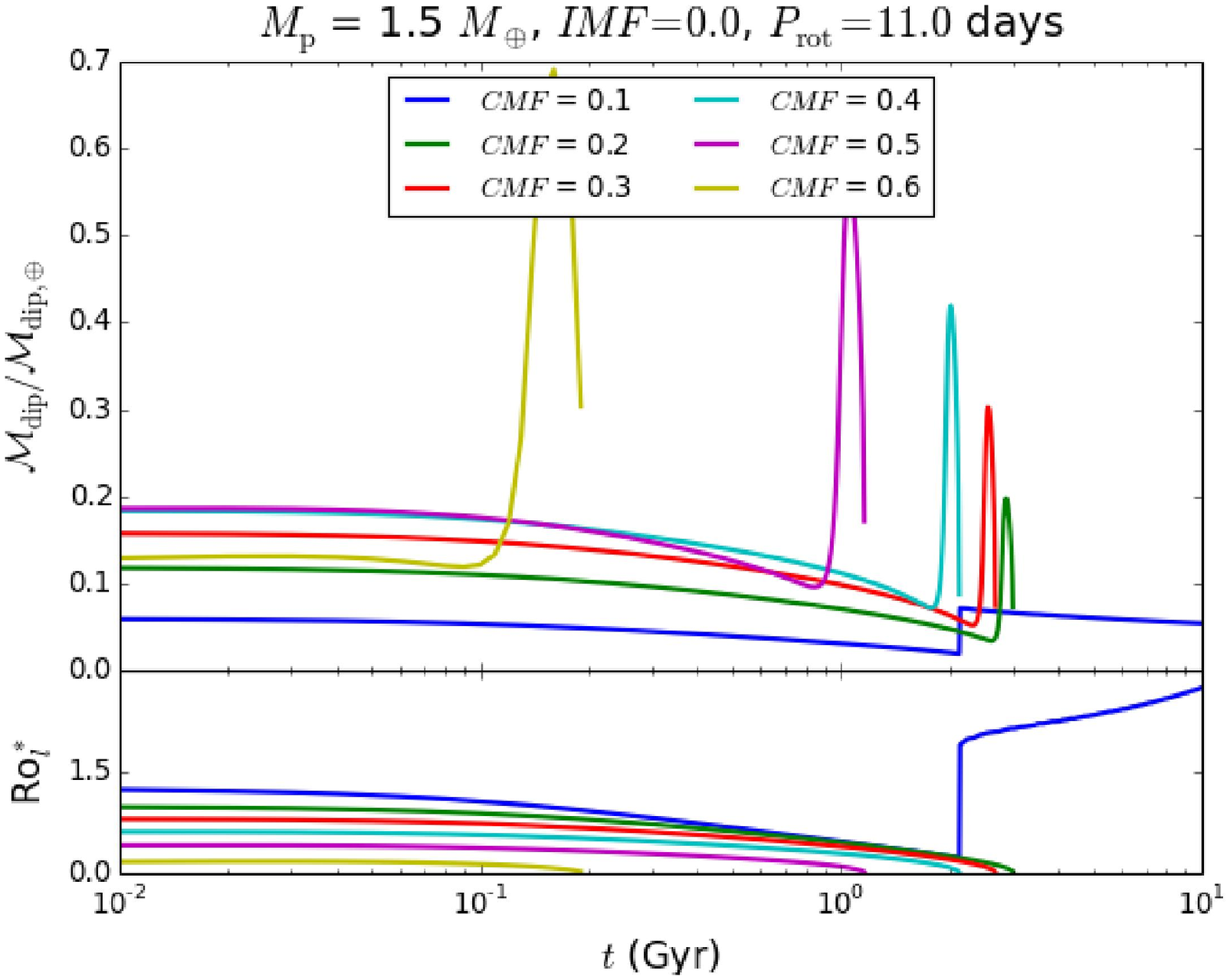}\hspace{0.5cm}\\\vspace{0.3cm}

   \includegraphics[width=70mm]{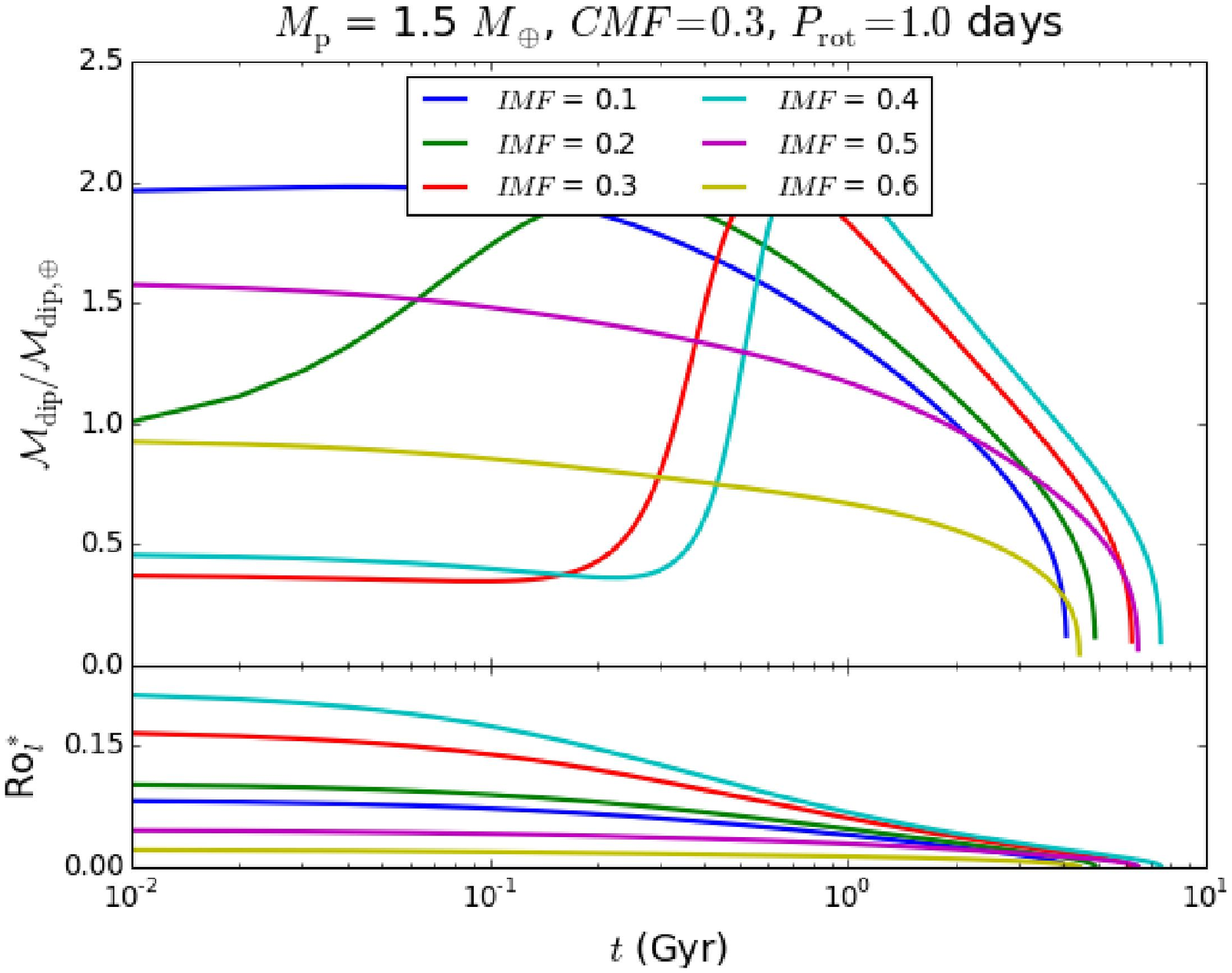}\hspace{0.5cm}
   \includegraphics[width=70mm]{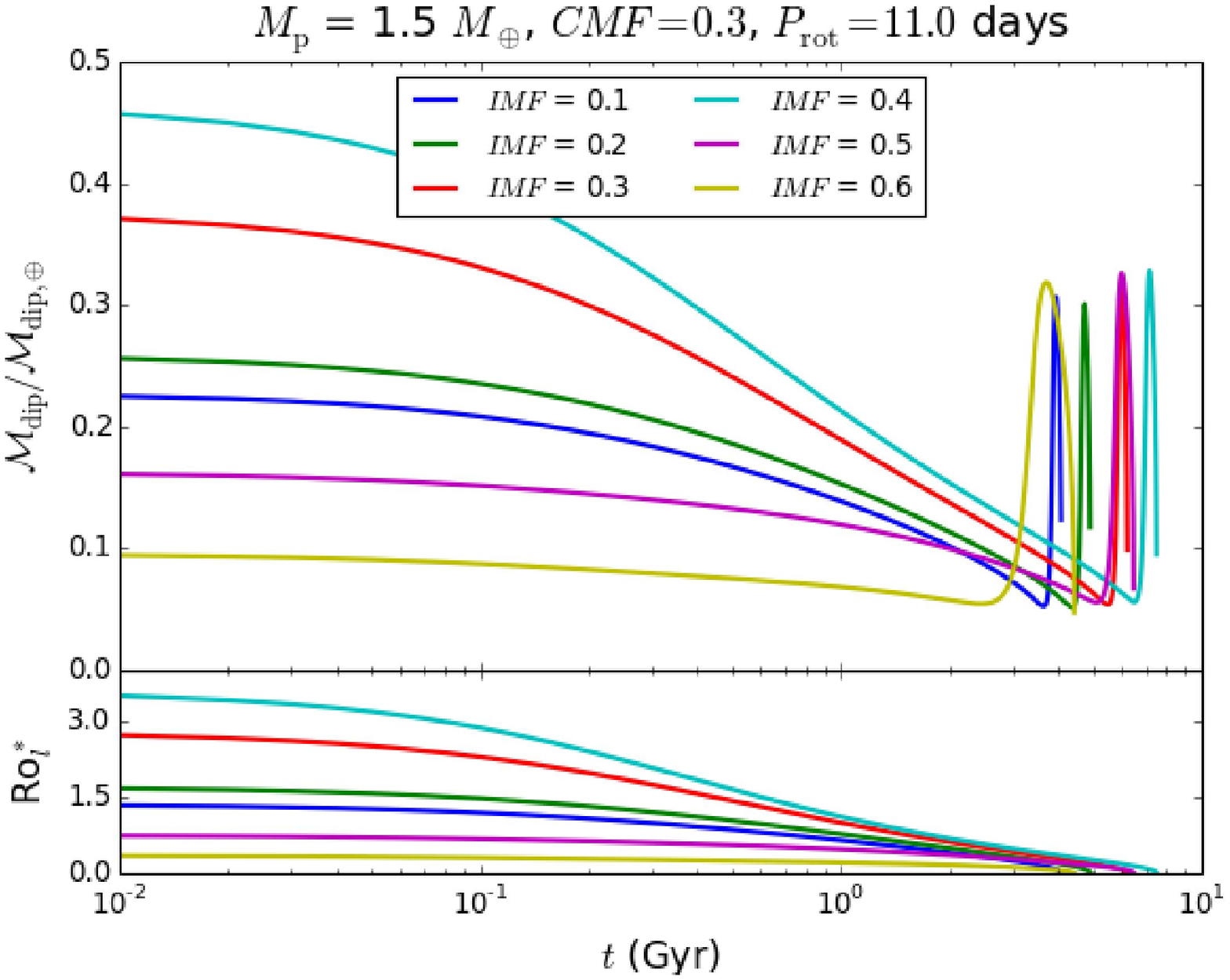}\hspace{0.5cm}\\\vspace{0.0cm}

  \scriptsize
  \caption{\hl{\hlll{Maximum} magnetic dipole moment and scaled local Rossby number (Equation \ref{eq:Rol}), calculated for a selected set of planetary models and two periods of rotations: 1 day (left column) and 11 days (right column).  Top row: Earth-like composition and different planetary masses.  Here CMF stands for Core Mass Fraction and IMF is the Ice Mass Fraction.  Middle row: dry planets ($IMF=0$) with fixed mass $M\sub{p}=1.5\,M_\oplus$ and different iron contents. Bottom row: planets with fixed mass $M\sub{p}=1.5\,M_\oplus$, Earth-like iron content $CMF=0.3$ and different water contents. \hlll{The spikes observed in the magnetic field intensity for the slowly rotating planets (right column), are a manifestation of the transition from a multipolar (Ro${}_l^*>0.1$) to a dipolar dominated magnetic field (Ro${}_l^*<0.1$).  This transition happens when the inertial convective forces in the iron core, become much lower than coriolis forces at later stages of thermal evolution.  Although this transition is more noticeable at low rotational rates, it is also observed in the case of fast rotating planets with a moderate water content (bottom row, left column)}}.\vspace{0.0cm}}
\label{fig:MagneticField}
\end{figure*}

\vspace{0.5cm}

\hl{In Figure \ref{fig:MagneticField} we show the \hlll{maximum} dipole moments calculated with the formulas described before \hlll{and in the case of the} planetary models of Figures \ref{fig:InteriorStructure}.  For comparison \hlll{purposes,} we have calculated the magnetic properties of the planets using two rotational periods: $P=1$ day (fast rotators) and $P=11$ days (slow rotators; \hlll{this is also the synchronous rotational period of \Pb, see \autoref{sec:Nominal}}).}

\hl{
The magnetic field strength for dry fast rotating planets evolves in a similar way as the available convective energy $Q\sub{conv}$ (Figure \ref{fig:ThermalEvolution}) as expected from the dynamo scaling law (Eq. \ref{eq:Brms}).  Water rich planets, however exhibits a curious behavior: in most cases the magnetic field is intensified at around half of their dynamo lifetime.   This intensification does not match any similar change in the available convective energy.  What we are seeing here is the transition from a multipolar \hlll{field} (low value of the dipolar component) to a \hlll{dipolar dominated one}.  This change is caused by the evolution of the Rossby number as the planet cools down.  As we explained before the available convective energy of water rich planets is relatively large.  As a consequence inertial forces driving convection in the outer core dominates over the coriolis forces for most part of the evolution.  In this condition, convection is turbulent and the magnetic field ``disorganized''.  When the planet has cool down enough, coriolis forces start to organize convection and a dipolar dominated magnetic field arises. The same behavior is observed in all the slowly rotating planets.  The dipolar dominated \hlll{field} in those cases, however, lasts for a small fraction of the dynamo lifetime.
}
\vspace{0.5cm}

\hl{Having a fully fledged planetary evolution model, we may now attempt to calculate the evolution of planets with properties (mass, composition and rotation) compatible with the observed characteristics of \Pb\ (planetary analogues).  We will focus here on to predict or constraint four basic properties: (1) planetary radius $R\sub{p}$, (2) average surface heat flux $q\sub{m}$, (3) dynamo lifetime $t\sub{dyn}$ and (4) minimum dipole moment, $\Mdip\sup{min}$. We will call these, the ``\gp'' of the planet.}

\section{\Pb\ analogues}
\label{sec:Analogues}

\subsection{Statistical strategy}
\label{subsec:StatisticalStrategy}

Only two properties of \Pb\ are presently known: its minimum mass $M\sub{min}=M\sub{p}\sin i=1.3\MEarth$ and orbital period $P\sub{orb}=11.3$ days \citep{AngladaEscude2016}.  Other key ``primary'' properties such as its actual mass $M\sub{p}$, bulk composition, planetary radius $R\sub{p}$,  rotational period $P\sub{rot}$, orbital inclination $i$ relative to the plane of the sky and eccentricity $e$ are yet unknown.

If we build \hl{a large synthetic sample} of \pa\  (\hl{a ``planetary ensemble''}), with each planet having random values of the unknown fundamental properties (which follow certain \hll{prior} distributions), \hl{and we use our model to compute the \gp\ of each planet in the sample,} we can estimate the \hl{posterior probability distribution} of the \gp\ in the ensemble. Using these estimated distributions we can predict \hl{or constrain} to a certain confidence level the expected \gp\ of \Pb\ and their analogues.

\begin{figure}
  \centering
   \includegraphics[width=0.4
  \textwidth]{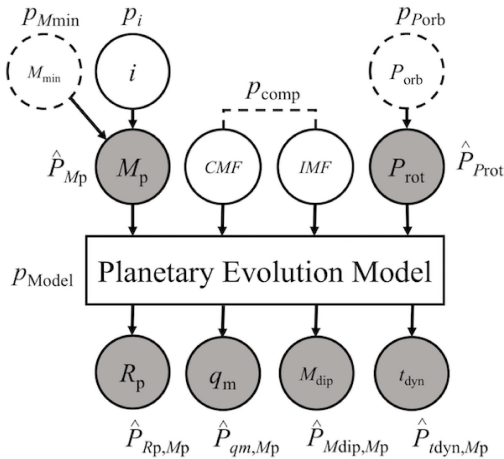}
  \scriptsize
  \caption{\hl{Statistical strategy.}}
\label{fig:StatisticalStrategy}
\end{figure}

\hl{A schematic representation of the statistical strategy devised here is presented in Figure \ref{fig:StatisticalStrategy}.  Each fundamental property (physical or observed) $X$, has an a \hll{prior} probability distribution $p_{\rm X}$.  Our purpose here is to estimate the posterior joint probabilities of the mass $M_{\rm p}$ and the corresponding magnetic property $Y$, namely $\hat{P}_{M{\rm p},Y}$.  Confidence level intervals for $Y$ are obtained by marginalizing ${\hat P}_{M{\rm p},Y}$ with respect to $M_{\rm p}$.  Our strategy resembles the Hierarchical Bayesian Models (HBM) which are becoming very popular in the context of exoplanetary sciences where the lack of information about the studied objects is the rule instead of the exception (see eg. \citealt{Chen2016} and references there in).}

\subsection{Model grid}
\label{subsec:ModelGrid}

For \hl{implementing the statistical strategy described in the previous section}, we build a grid of planetary models with masses ranging from 1 to 7 $\MEarth$ (resolution of 0.3-0.5$\MEarth$).  This set of masses not only include the minimum mass of \Pb, but cover all possible orbital inclinations between \hl{$10^\circ-90^\circ$}.  \hlll{Though inclinations $i\lesssim 10^\circ$ ($M\sub{p}>7\,\MEarth$) are in principle possible, planets that massive would produce detectable astrometric signals in Proxima Centauri \citep{Barnes2016}. Moreover, even accepting very low inclinations, more than 85\% of the analogues would have masses below our model maximum of $7\,\MEarth$ (see below)}.

For each mass \hl{in the grid}, our codes are run for different core, ice and mantle mass-fractions (CMF, IMF and MMF respectively). In our \hlll{model} CMF ranges from 0.1 to 0.8 (0.1 steps) and IMF \hl{goes} from 0.0 to 0.9-CMF.  Pure iron, pure rock or pure ice analogues were not considered.

\hl{A} total \hl{of} 1,100 different combinations of mass and compositions were simulated. \hl{The} grid not only encloses the properties of \Pb, but also the properties of many other discovered and yet to be discovered earth-like exoplanets.

\subsection{Planetary ensamble}
\label{subsec:PlanetaryEnsamble}

To build our ensemble of \Pb\ analogues only three of the fundamental properties, \hlll{namely}, inclination $i$, rotational period $P\sub{rot}$ and bulk composition (CMF,IMF) are randomly generated. \hl{For simplicity, we have assumed that the observable quantities $M\sub{min}$ and $P\sub{orb}$ are precisely known, ie. $p_{M{\rm min}}$ is a discrete probability distribution that is equal to 1 for the observed value $M\sub{min}$ and 0 otherwise.  The same \hlll{assumption is applied to} $p_{P{\rm orb}}$.}  In all cases, eccentricity is assumed close to 0 (see Section \ref{sec:Discussion} for a discussion of this assumption).

Inclinations are generated assuming a flat prior in the interval $i\in[10^\circ,90^\circ]$, \hl{ie. $p\sub{i}={\cal U}(10^\circ,90^\circ)$}.  Compositions (CMF,IMF) are generated in such a way that every possible combination in our grid are equally probable (flat composition prior \hl{$p\sub{comp}$}).

The period of rotation is a key quantity at determining the \gp.  Since the planet is close to its hots star (semi-major axis $a=0.05 AU$), its period of rotation should be in a resonance with its orbital period \citep{Makarov2013b}. The most obvious resonance is the synchronous rotation (1:1).  However, even if the orbital eccentricity is small, it could be trapped in a supersynchronous resonance, ie. 3:2 or 2:1 \citep{Makarov2013a,Makarov2013b,Ribas2016}. The probability of being trapped in those rotational states depends not only on eccentricity, but also on triaxiality \citep{Ribas2016}, bulk composition \citep{Cuartas2016} and detailed rheological properties \citep{Makarov2013a}.  In the absence of any prior knowledge about these parameters, we will assume that \Pba\ have, with equal probability, one of two rotational periods: 11.3 days (syncrhonous rotation) or 7.5 days (3:2 resonance).  \hl{Equivalently the posterior distribution function $\hat{P}_{P{\rm rot}}$ will be assumed discrete, such that $\hat{P}_{P{\rm rot}}(11.3)=1/2$ and $\hat{P}_{P{\rm rot}}(7.5)=1/2$}.


\vspace{-0.5cm}
\section{Magnetic properties of \Pb\ analogues}
\label{sec:Results}

\subsection{Nominal analogues}
\label{sec:Nominal}

We call ``\na'' those planets having the same observed minimum mass of \Pb\ and a rotational period equal to 11.3 days (synchronous rotation). \hl{In statistical terms \na\ are those obtained with the priors $\hat{P}_{M{\rm p}}(M\sub{min})=1$ and $\hat{P}_{P{\rm rot}}(11.3)=1$}.

\begin{figure}
  \centering
  \vspace{0.2cm}
   \includegraphics[width=65mm]{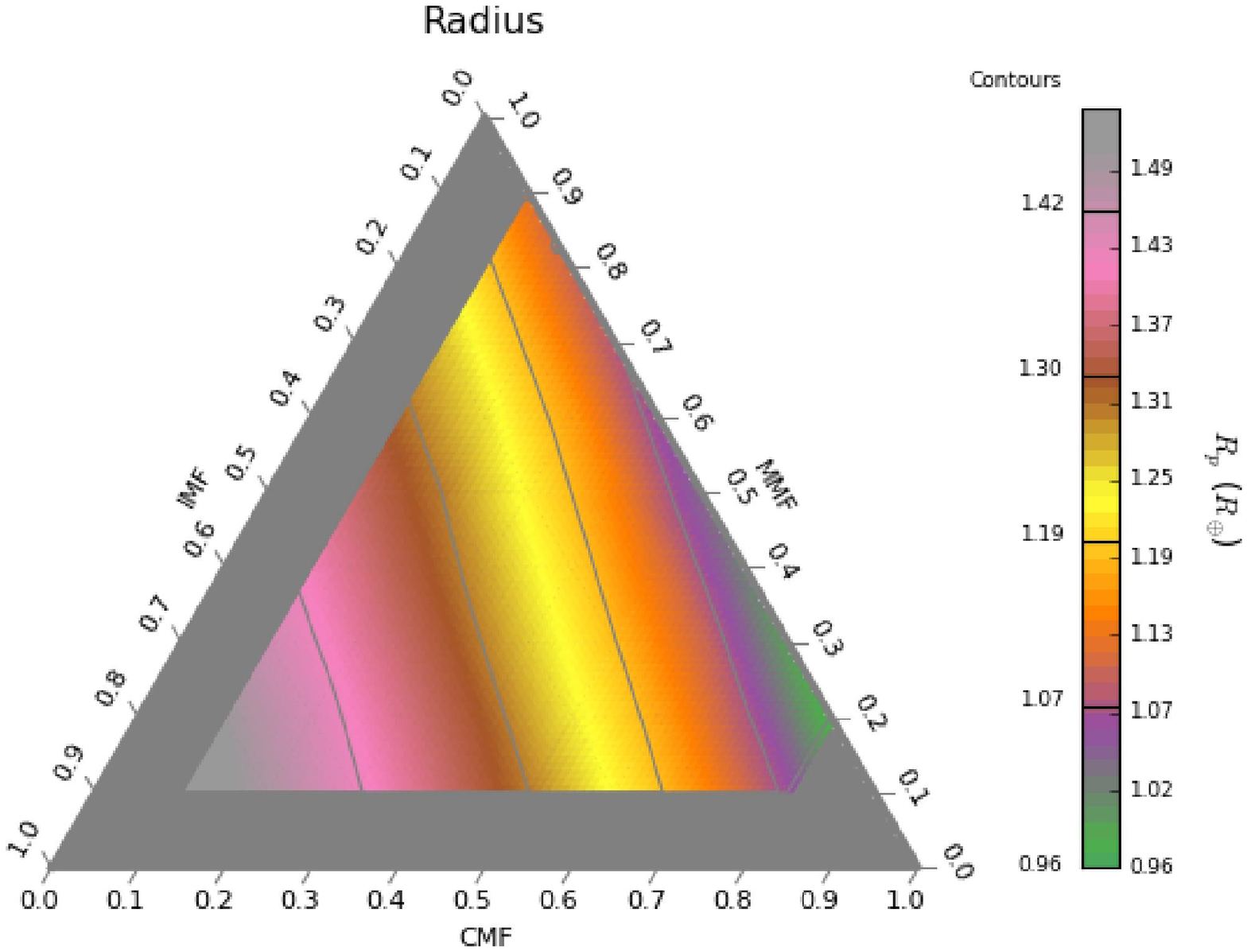}\\
   \includegraphics[width=65mm]{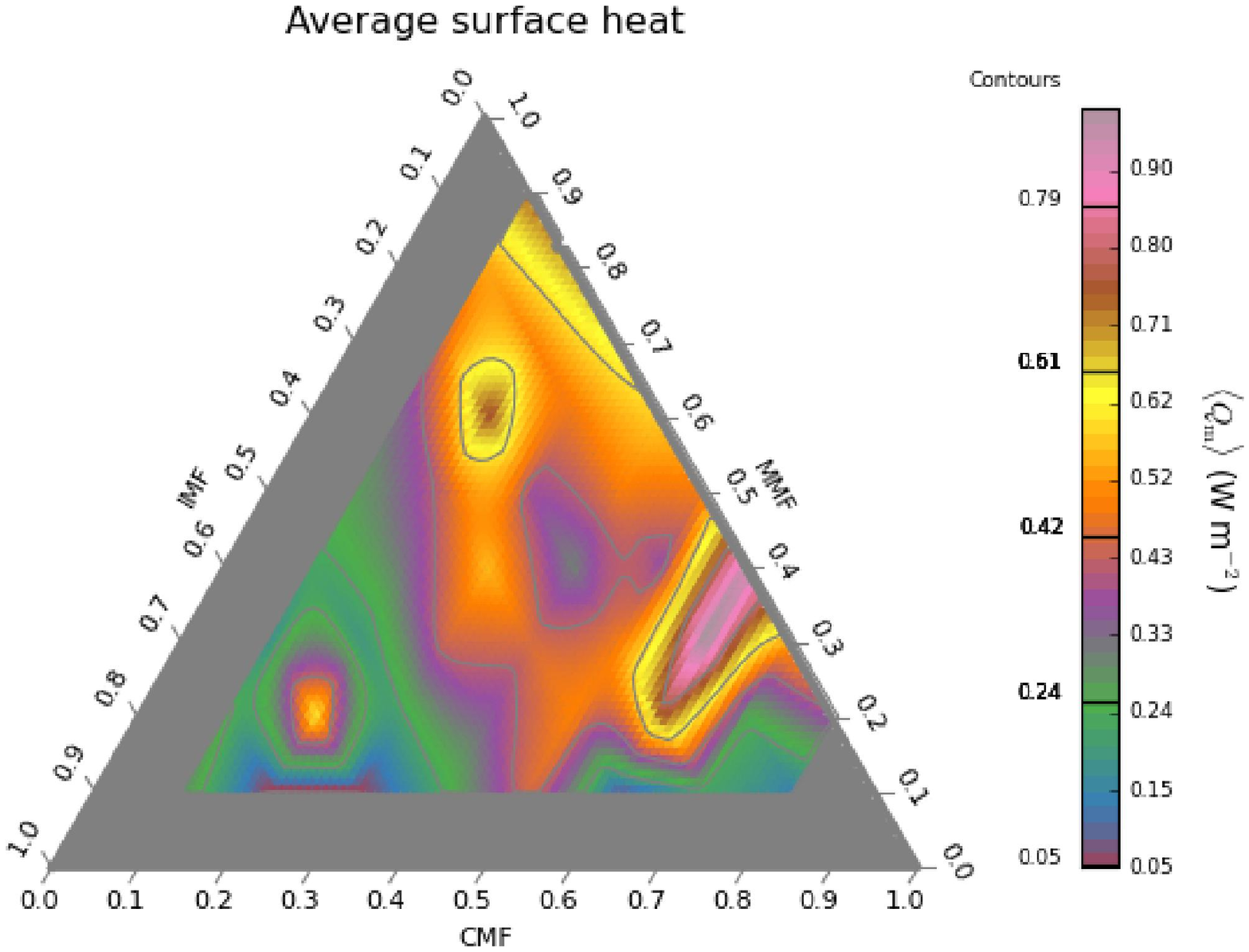}\\
   \includegraphics[width=65mm]{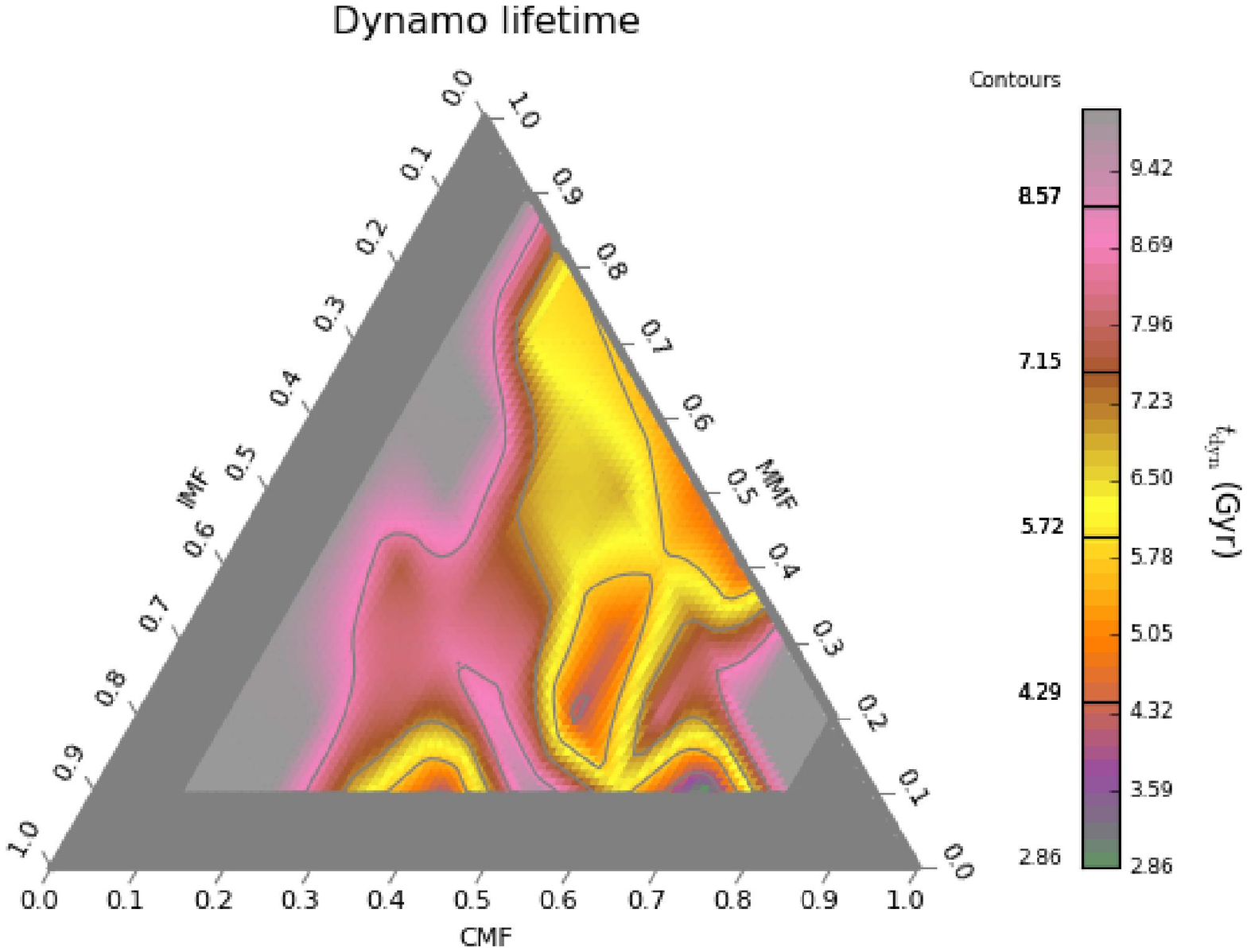}\\
   \includegraphics[width=65mm]{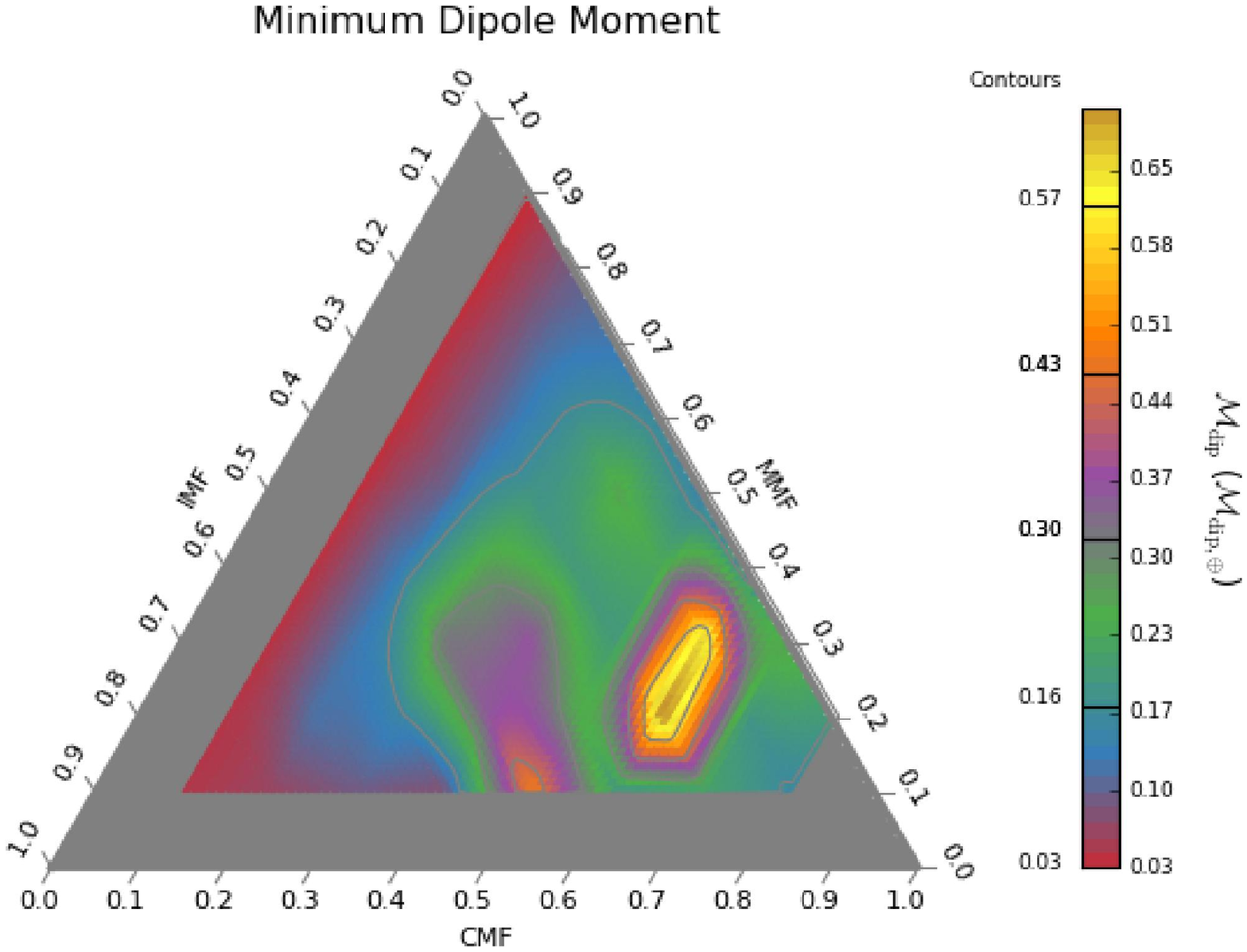}
  \scriptsize
  \caption{Ternary diagrams of the \gp\ for nominal \Pba. Gray areas correspond to core-less or mantle-less objects where no planetary model were calculated.\vspace{0.0cm}}
\label{fig:NominalProximab}
\end{figure}

\begin{figure*}
  \centering
  \vspace{0.2cm}
   \includegraphics[width=70mm]{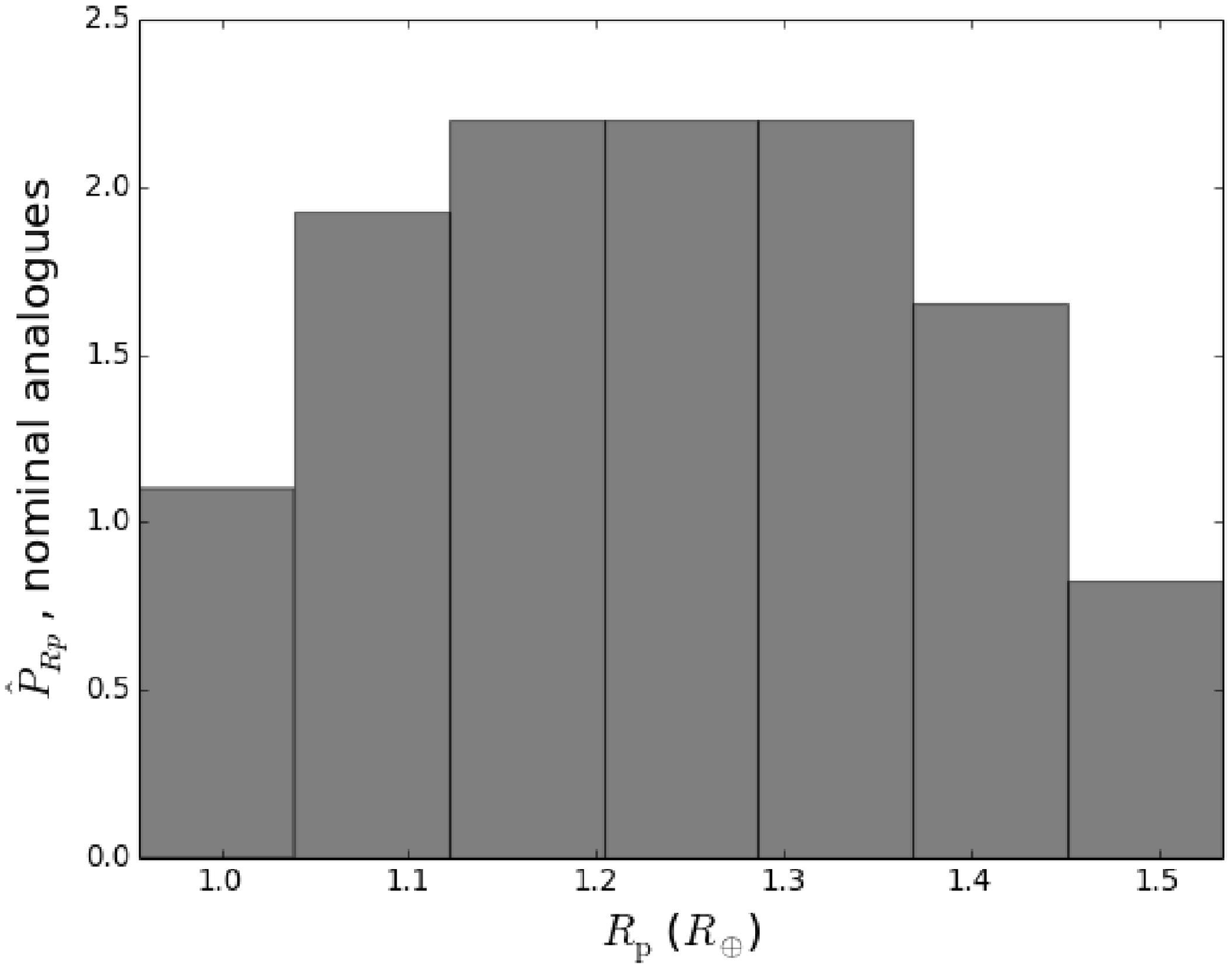}\hspace{0.5cm}
   \includegraphics[width=70mm]{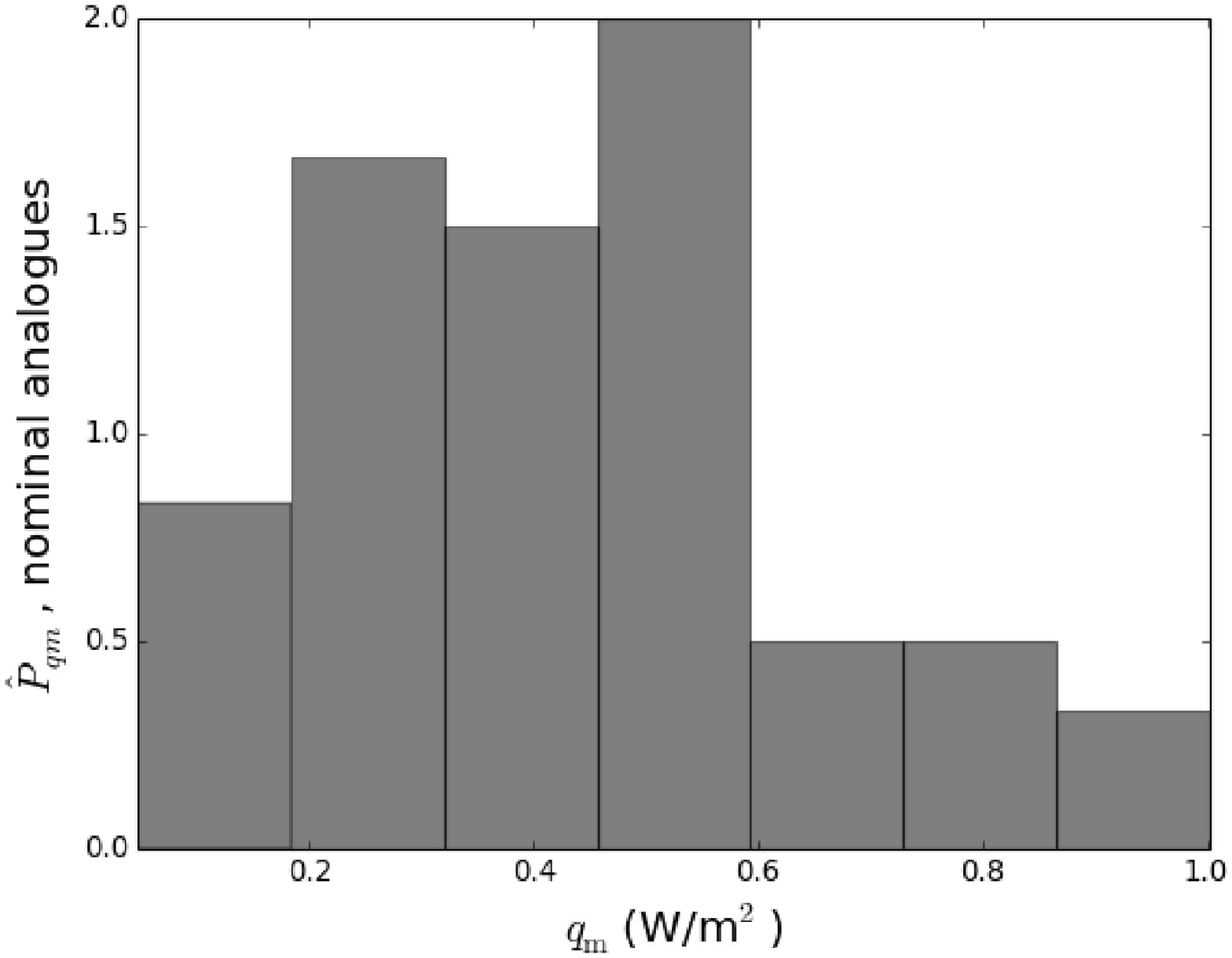}\\\vspace{0.5cm}
   \includegraphics[width=70mm]{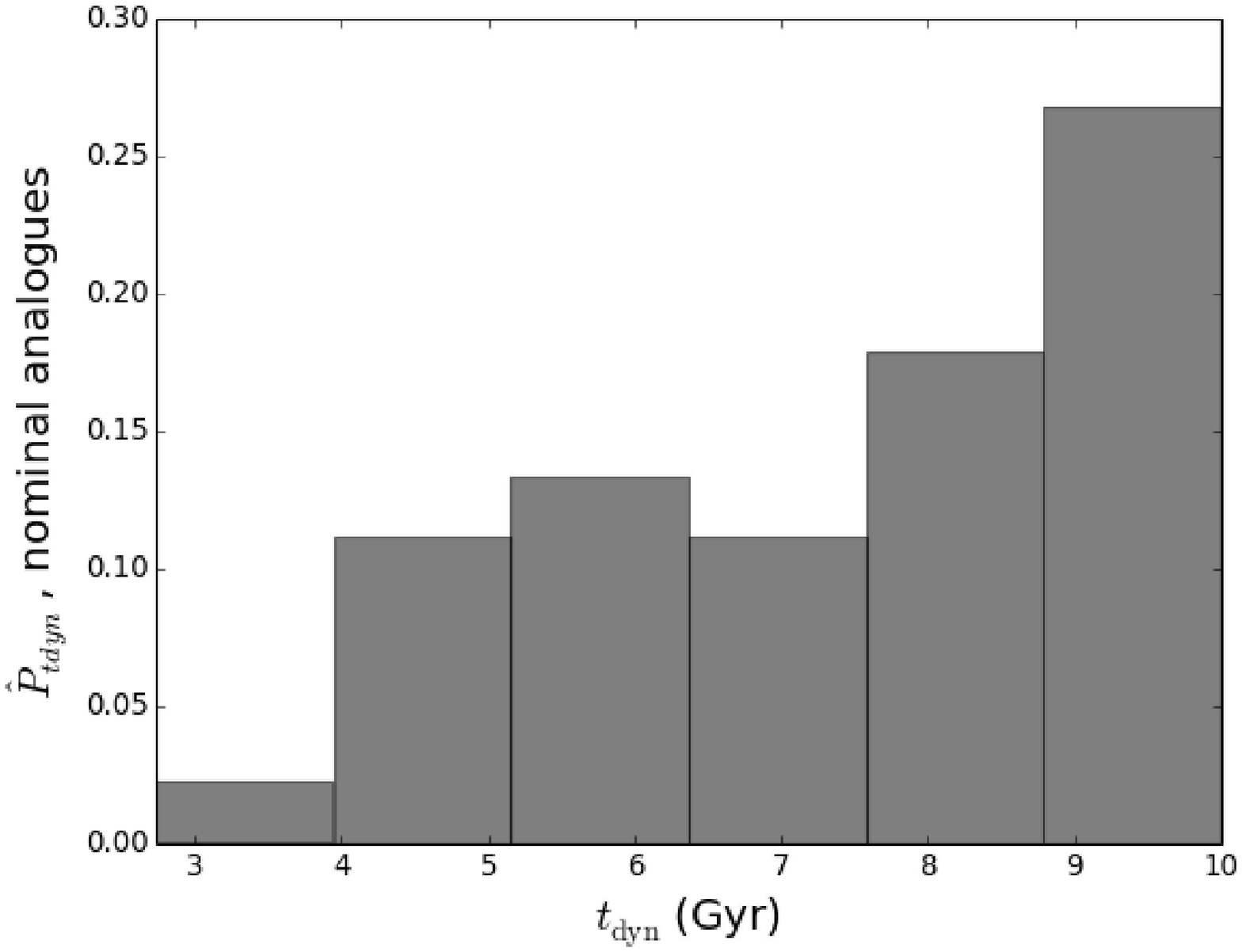}\hspace{0.5cm}
   \includegraphics[width=70mm]{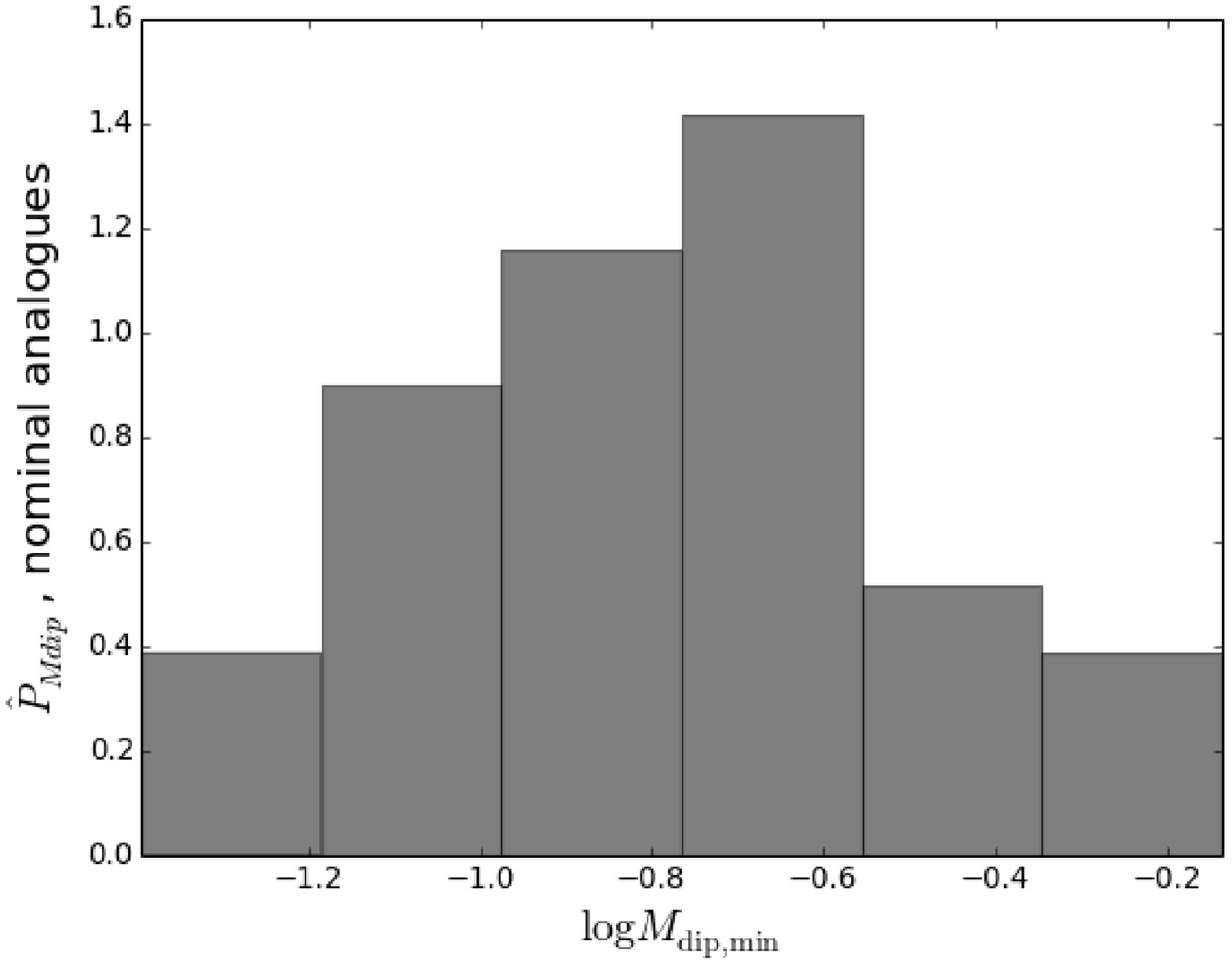}
  \scriptsize
  \caption{Posterior distribution of the \gp\ of the \na.\vspace{0.0cm}}
\label{fig:NominalPosterior}
\end{figure*}

In Figure \ref{fig:NominalProximab} we show contour plots in ternary diagrams of the 4 key \gp\ of these analogues. \hl{Since we are assuming flat priors for compositions, ie. every composition in our grid is equally probably, the posterior probability distribution of each \gp\ in the case of \na\ can be estimated with the frequency of the properties calculated in all grid points. In Figure \ref{fig:NominalPosterior} we plot the resulting estimated posterior distributions.}

The nominal \Pba\ have radii in the range $0.96-1.53 \REarth$ with a median value of 1.24 $\REarth$. More precisely:

\hl{
$$
R\sub{p}\sup{nom}=1.24\pm 0.16\,\REarth\,(70\%\,{\rm CL})
$$
}

Most of the uncertainty in radius comes from the uncertainty in the IMF.  If for instance we assume a negligible IMF, the radius of the nominal analogues is better constrained:

\hl{
$$
R\sub{p}\sup{nom,dry}=1.04\pm 0.06 \REarth\,(70\%\,\mbox{CL})
$$
}

This value is in agreement with the estimations by \citet{Kipping2016} using the novel ``forecasting'' technique recently introduced by \citet{Chen2016}.

We identify a ``sweet-spot'' around CMF$\approx$0.6, MMF$\approx$0.35 and IMF$\approx$0.15 \hl{where favorable magnetic conditions could arise}.  These ``iron-rich'' ocean analogues, develop long-lived dynamos and intense magnetic fields.  Despite having a rotational rate $\sim$10 times slower than Earth, these \na\ are magnetically alike to Earth.

For most compositions, \Pb\ {\na} have dynamo lifetimes larger than planetary age, $\tau_\star=4.8\pm 1$ Gyr \citep{Bazot2016,Barnes2016,Thevenin2002,Thoul2003}.  Only in 10\% of the compositions (eg.
IMF$\approx$0, CMF=MMF$\approx$ 0.5), the dynamo has been already shut down or it is disappearing.

\hl{
According to the posterior distribution of the minimum dipole moment  (see Figure \ref{fig:NominalPosterior}) the magnetic field strength of \Pb\ \na\ is:
}

\hl{
\begin{eqnarray}
\nonumber
\log(\Mdip\sup{nom}) & = & -0.77^{+0.27}_{-0.26}\,(70\%\,\mbox{CL})\\
\nonumber
\Mdip\sup{nom} & = & 0.17^{\times 1.86}_{\div 1.86}\;\MdipEarth\,(90\%\,\mbox{CL})
\end{eqnarray}
}

Equivalently,

\hl{
\begin{eqnarray}
\nonumber
\log(B\sub{dip}\sup{nom}) & = & 0.67^{+0.44}_{-0.19}\,(70\%\,\mbox{CL})\\
\nonumber
B\sub{dip}\sup{nom} & = & 4.7^{\times 2.8}_{\div 1.5}\;\mu\mbox{T}\,(70\%\,\mbox{CL})
\end{eqnarray}
}

\vspace{-0.5cm}
\subsection{Planetary ensemble}

\begin{figure*}
  \centering
  \vspace{0.2cm}
   \includegraphics[width=100mm]{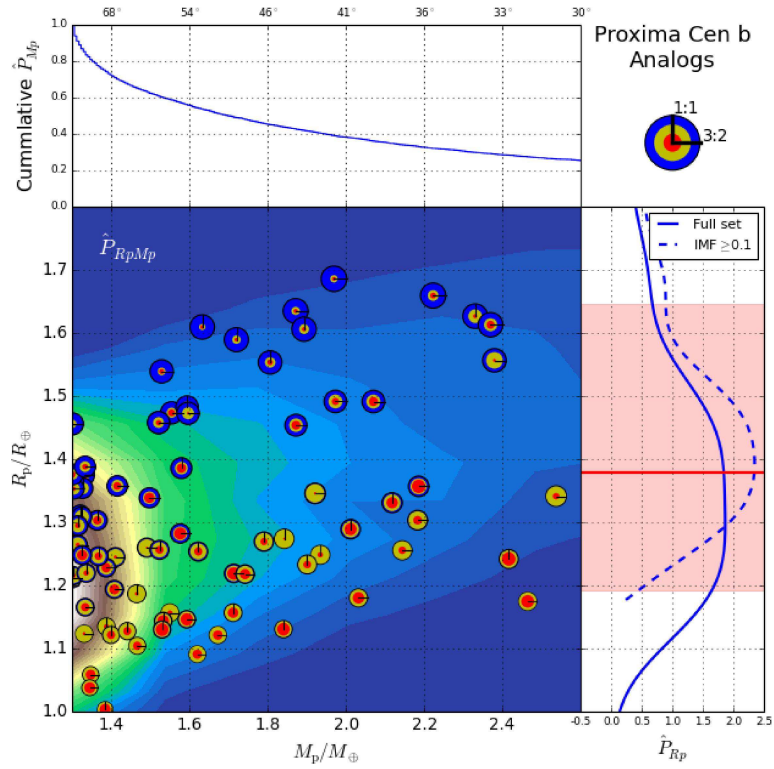}\hspace{0.9cm}\vspace{0.0cm}
  \scriptsize
  \caption{Mass-radius diagram (central panel), cumulative marginal distribution of planetary mass (top inset) and marginal posterior distribution of planetary radii (right inset) for \Pba.  The joint posterior distribution $\hat{P}_{RpMp}$ is represented by the background contour plot.  Layered circles illustrate the properties of 100 selected planets. The radius of each circle is proportional to planetary radius.  The width of the color layers are proportional to mass-fractions: iron core (red), rocky mantle (yellow) and ice crust (blue). Periods of rotation are represented with a vertical (1:1 resonance) or a horizontal (3:2 resonance) black solid line.\vspace{0.1cm}}
\label{fig:MassRadius}
\end{figure*}

In Figure \ref{fig:MassRadius} we summarize in a mass-radius diagram the most important features of the planetary ensemble of \Pb\  analogues. For illustration purposes, we show in this figure a selected subset of the planets (colored circles). Size and colors of the layers in the circles represent mass fraction of ice, silicates and iron. Contours in the main panel of Figure \ref{fig:MassRadius} represent the joint posterior probability distribution function $\hat{P}_{RpMp}$.

In the upper and right panels, the estimated marginal cumulative distribution of $M_p$, namely $\int_{M_p}^{\infty} \hat{P}_{Mp} dM_p$, and the marginal posterior distributions of planetary radius $\hat{P}_{Rp}$ are also shown.

\hl{A uniform prior in the inclination $i$ produce a highly concentrated planetary mass posterior distribution.  Accordingly, $\sim 80\%$ of the \pa\ have masses below 2.6 $M_\oplus$.}

\hl{From the estimated marginal posterior distribution of planetary radius $\hat{P}_{Rp}$ in the right panel of Figure \ref{fig:MassRadius}} we conclude that provided uniform priors for composition and orbital inclination, the planetary radius of \Pb\ analogues will be:

$$
R\sub{p}=1.38^{+0.26}_{-0.19}\REarth\,(70\%\,\mbox{CL})
$$.

If on the other hand we \hl{only include} planets having significant amount of volatiles (hereafter ``ocean planets'', IMF$\geq$0.2), the resulting marginalized posterior distribution \hl{(dashed curve in the right panel of Figure \ref{fig:MassRadius})} gives us a slightly larger radius:

$$
R\sup{ocean}\sub{p}=1.46^{+0.29}_{-0.15}\REarth\,(70\%\,\mbox{CL})
$$

\begin{figure*}
  \centering
  \vspace{0.0cm}
  \includegraphics[width=100mm]{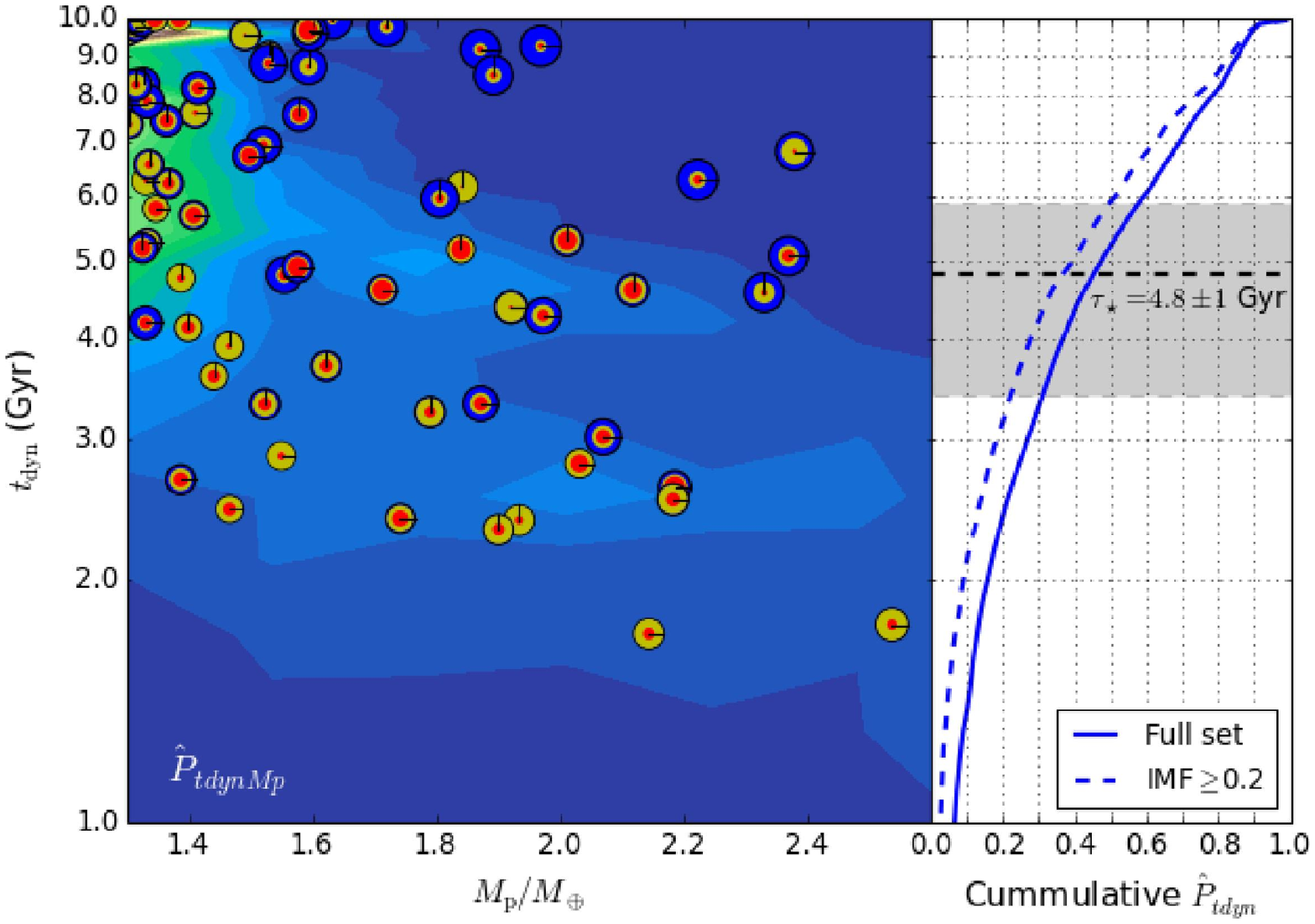}
  \includegraphics[width=100mm]{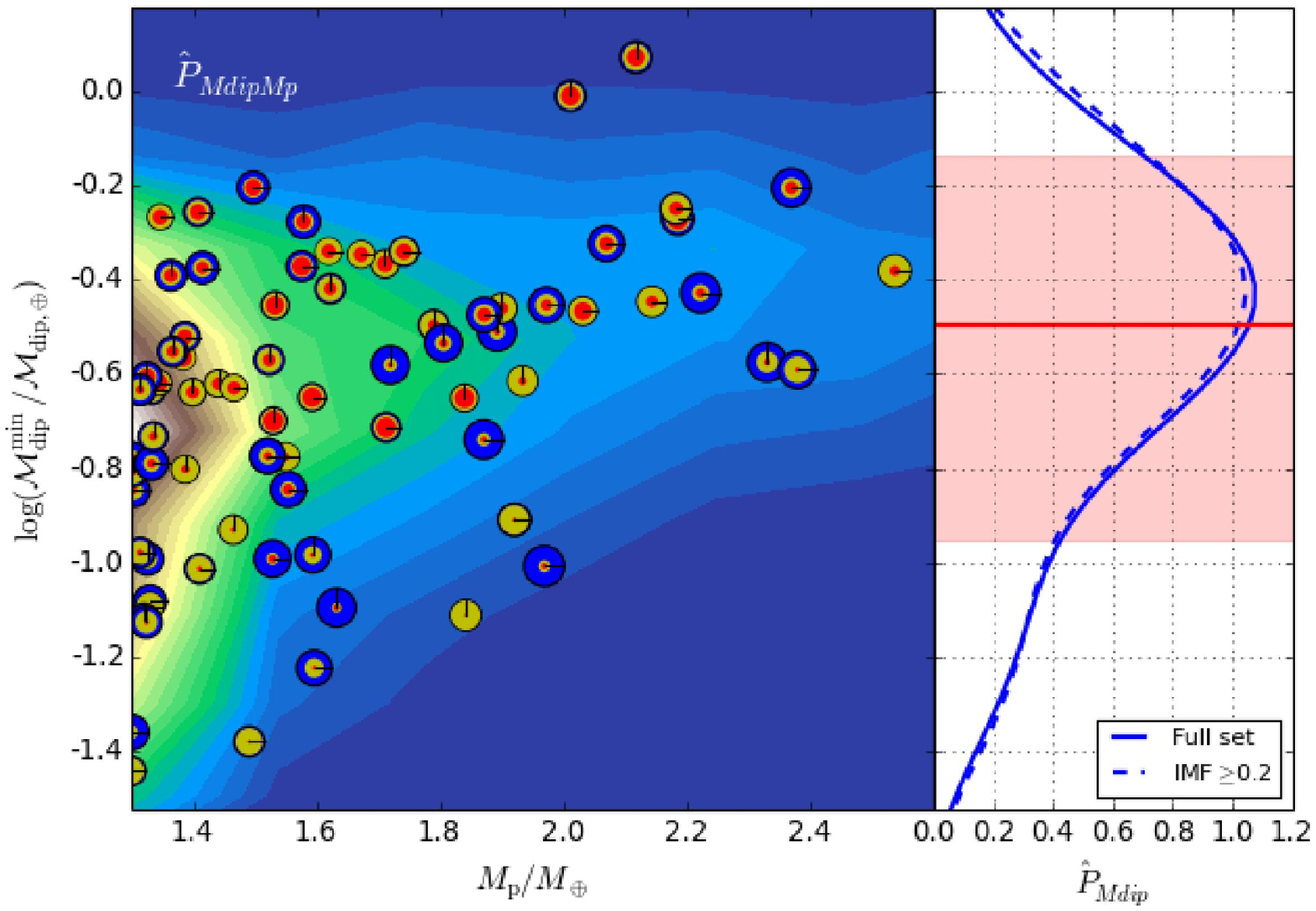}\\\vspace{0.0cm}
  \scriptsize
  \caption{Posterior joint and marginal distributions of dynamo life-time (left panel) and minimum dipole moment (right panel) for \Pb\ analogues.\vspace{0.0cm}}
  \label{fig:magnetic}
\end{figure*}

A similar analysis was performed for the case of the dynamo lifetime and minimum dipole moment. Results are shown in Figure \ref{fig:magnetic}.

As in the nominal case, most \pa\ have long-lived dynamos.  \hl{For the particular case of \Pb} we notice that most ($\gtrsim 60-70\%$) of the analogues have dynamo lifetimes $\gtrsim \tau\sub{\star}$ \hl{(gray shaded area in the inset panel of Figure \ref{fig:magnetic})}.  However, If we restrict the ensemble to \hl{ocean planets IMF$\geq$0.2, the fraction of analogues having long-lived dynamos increases in almost 10$\%$.  This fact is consistent with the results in Figure \ref{fig:MagneticField}.}

We find that a non-negligible fraction of analogues, mostly iron-rich planets with masses larger than 2$\MEarth$, despite of having large rotational periods, have also magnetic fields as intense or even stronger than Earth.  Comparing the dynamo life-time and magnetic dipole moment joint posterior distributions, we realize that \hl{usually the same} planets having stronger fields are also those with shorter-lived dynamos. \hl{This is consistent with the result in the upper row of Figure \ref{fig:MagneticField}}.

It is interesting to notice that the posterior distribution of minimum dipole moments \hl{(right panel of Figure \ref{fig:magnetic})} is not significantly modified when we exclude volatile rich planets.  This would imply that for the sake of estimating sample average magnetic field strength, bulk composition is not as important as in the case of other \gp.

\hl{Using the posterior distribution in Figure \ref{fig:magnetic}} we predict that the minimum dipole moment for \Pb\ analogues is:

$$
\log(\Mdip/\MdipEarth)=-0.49^{+0.36}_{-0.47}\,(70\%\,\mbox{CL})
$$,

or equivalently:

$$
\Mdip=0.32^{\times 2.3}_{\div 2.9}\;\MdipEarth\,(70\%\,\mbox{CL})
$$

\hl{Correspondingly,} the surface magnetic field strength will be:

$$
B_{\rm dip}=5.4^{\times 2.2}_{\div 3.9}\;\mu{\rm T}\,(70\%\,\mbox{CL})
$$

\vspace{-0.5cm}
\section{Discussion and conclusions}
\label{sec:Discussion}

The results presented in this paper rely on relatively robust models of planetary interior structure, thermal evolution and dynamo scaling laws. The discovery of \Pb\ is an unique opportunity to compare the predictions of these models with an exoplanet that could be observed with unprecedented detail in the near future and, hopefully, reached by space probes in the coming centuries.

\hll{We have used in our interior structure and thermal evolution models rather standard values of the rheological and thermodynamic properties of the materials inside the planet (thermal conductivities, gruneisen parameter, thermal expansivities, viscosity parameters, etc.).  Many of these parameters are still very uncertain, especially in the case of planets more massive than Earth. Some of them depend on composition and even may change as the planet evolve. As a consequence the value that we obtained here for the magnetic properties would be different if better information about these properties is obtained from laboratory experiments or constraint from future observations.}

\hll{In our models we have assumed the same composition for all the planetary cores despite the differences in mantle and water envelope composition.  It should be stressed that assuming different iron alloys or the presence of other elements such as sulfur, could significantly modify the results as it has turned to be the case of Mercury in the Solar System (see eg. \citealt{Harder2001}).}

Other models of planetary thermal and magnetic evolution have appeared in literature since \citet{Zuluaga2013}.  Of particular relevance is the recent model by \citet{Driscoll2015} that includes the effects of tidal heating.  If tidal heating is negligible (which is the case for a close-in low eccentric planet trapped in synchronous rotation), both models \hll{should have similar predictions (under the same assumptions)}; their results will not differ more than \hl{the} discrepancies arising from the very uncertain physical parameters on which \hl{these} models depends \hl{on}.

If future observations confirm the existence of another planet in the system, as it was suggested by \citet{AngladaEscude2016}, or are able to measure a non negligible orbital eccentricity ($e>0.06$) we will need to update these results by including the effects of tidal heating.  Meanwhile, however, including these effects (that require at least another 2 or 3 unknown input free parameters) would just reduce the robustness of our predictions.

Papers published so far about \Pb\ \citep{Ribas2016,Barnes2016,Coleman2016,Turbet2016} have focus on modeling a lot of detailed properties of a planet with 1.3 $\MEarth$ and compositions similar to that of Earth.  While this paper has not been as ambitious and exhaustive as those other works, we have approached to the problem in a fundamentally different way.  We built a grid of planetary models, having a large range of masses and compositions, and calculated posterior probability distributions of a few key properties.

Although the statistical significance of \Pb\ discovery is impressive (FAP$\sim 10^{-7}$, $B_1/B_0\sim 10^7$, \citealt{AngladaEscude2016}) we should still seek for an independent confirmation.  Even in the very unlikely case that the signal of \Pb\ have another origin (see a recent example in \citealt{Anglada2015,Hatzes2016}), our results are still valid for confirmed planets having similar basic properties (minimum mass, orbital period and stellar mass).

Our models have several important caveats that could make their numerical predictions biased or simply wrong.

We have completely neglected orbital eccentricity.  Although \citet{AngladaEscude2016} observations are compatible with eccentricities as large as 0.35, it does not directly suggest that the planet has an eccentric orbit.  Still, the dynamics of the triple system or the stellar stream to which the host star belongs, suggest that past perturbations could have excited a non-zero eccentricity \citep{Barnes2016}.

The presence of another super-Earth in the system is neither confirmed nor discarded by observations.  If a second planet exist, the orbit of \Pb\ could be perturbed and have an oscillating non-negligible eccentricity.

We assumed that the abundance of key radionuclides, such Th-232, U-238 and K-40, in the mantle of all our planetary analogues is identical to that of Earth.  However, it has been shown the abundance of Eu (a proxy for r-process elements such as Th and U) in the atmosphere of $\alpha$ Cen A is lower than that of the Sun \citep{Hinkel2013}, suggesting that \Pb\ could have also a lower content of heavy radionuclides.  On the other hand, volatile-rich analogues of \Pb\ could also have larger contents of K-40 (a volatile element), provided they form beyond the snow-line.  The effect of a larger radiogenic inventory has been studied by \citet{Barnes2016} (only in the case of planets having the minimum mass).  Larger radionuclides abundances delay the solidification of an inner-core but increase the core convective power. In our case if the abundance of Th and U is relatively low, the predicted magnetic field strength will be slightly lower.  However, if our ocean planets have a significant abundance of K-40, they could lack entirely of a dynamo.

\hll{The small distance of Proxima Cen b opens a new realm of possibilities for the observation of the effects that a
planetary magnetic field has around the planet.  Synchrotron
emission and auroral ultraviolet radiation could be detected if we can resolve the planet with future telescopes and radio telescopes. Moreover, if a space probe is designed and launched in the next decades we can measure in situ the magnetic properties of the planet in a similar fashion as modern Solar System probes do with planets and satellites.}

\vspace{0.5cm}

We have used NASA ADS Bibliographic Services. Most of the computations that made possible this work were performed with {\tt Python 2.7} and their related tools and libraries, {\tt iPython} \citep{Perez2007}, {\tt Matplotlib} \citep{Hunter2007}, {\tt scipy} and {\tt numpy} \citep{Van2011}. This work is supported by Vicerrectoria de Docencia-UdeA and the {\it Estrategia de Sostenibilidad 2014-2015 de la Universidad de Antioquia}.  SB acknowledges support from the International Max-Planck Research School for Astronomy and Cosmic Physics at the University of Heidelberg (IMPRS-HD) and financial support the Deutscher Akademischer Austauschdienst (DAAD) through the program Research Grants - Doctoral Programmes in Germany (57129429).


\begin{thebibliography}{53}
\expandafter\ifx\csname natexlab\endcsname\relax\def\natexlab#1{#1}\fi
\providecommand{\url}[1]{\texttt{#1}}
\providecommand{\href}[2]{#2}
\providecommand{\path}[1]{#1}
\providecommand{\DOIprefix}{doi:}
\providecommand{\ArXivprefix}{arXiv:}
\providecommand{\URLprefix}{URL: }
\providecommand{\Pubmedprefix}{pmid:}
\providecommand{\doi}[1]{\href{http://dx.doi.org/#1}{\path{#1}}}
\providecommand{\Pubmed}[1]{\href{pmid:#1}{\path{#1}}}
\providecommand{\bibinfo}[2]{#2}
\ifx\xfnm\relax \def\xfnm[#1]{\unskip,\space#1}\fi
\bibitem[{Agostini et~al.(2015)Agostini, Appel, Bellini, Benziger, Bick,
  Bonfini, Bravo, Caccianiga, Calaprice, Caminata et~al.}]{Agostini2015}
\bibinfo{author}{Agostini, M.}, \bibinfo{author}{Appel, S.},
  \bibinfo{author}{Bellini, G.}, \bibinfo{author}{Benziger, J.},
  \bibinfo{author}{Bick, D.}, \bibinfo{author}{Bonfini, G.},
  \bibinfo{author}{Bravo, D.}, \bibinfo{author}{Caccianiga, B.},
  \bibinfo{author}{Calaprice, F.}, \bibinfo{author}{Caminata, A.}, et~al.,
  \bibinfo{year}{2015}.
\newblock \bibinfo{title}{Spectroscopy of geoneutrinos from 2056 days of
  borexino data}.
\newblock \bibinfo{journal}{Physical Review D} \bibinfo{volume}{92},
  \bibinfo{pages}{031101}.
\bibitem[{Anglada-Escud{\'e} et~al.(2016)Anglada-Escud{\'e}, Amado, Barnes,
  Berdi{\~n}as, Butler, Coleman, de~la Cueva, Dreizler, Endl, Giesers, Jeffers,
  Jenkins, Jones, Kiraga, K{\"u}rster, L{\'o}pez-Gonz{\'a}lez, Marvin, Morales,
  Morin, Nelson, Ortiz, Ofir, Paardekooper, Reiners, Rodr{\'\i}guez,
  Rodrίguez-L{\'o}pez, Sarmiento, Strachan, Tsapras, Tuomi and
  Zechmeister}]{AngladaEscude2016}
\bibinfo{author}{Anglada-Escud{\'e}, G.}, \bibinfo{author}{Amado, P.J.},
  \bibinfo{author}{Barnes, J.}, \bibinfo{author}{Berdi{\~n}as, Z.M.},
  \bibinfo{author}{Butler, R.P.}, \bibinfo{author}{Coleman, G.A.L.},
  \bibinfo{author}{de~la Cueva, I.}, \bibinfo{author}{Dreizler, S.},
  \bibinfo{author}{Endl, M.}, \bibinfo{author}{Giesers, B.},
  \bibinfo{author}{Jeffers, S.V.}, \bibinfo{author}{Jenkins, J.S.},
  \bibinfo{author}{Jones, H.R.A.}, \bibinfo{author}{Kiraga, M.},
  \bibinfo{author}{K{\"u}rster, M.}, \bibinfo{author}{L{\'o}pez-Gonz{\'a}lez,
  M.J.}, \bibinfo{author}{Marvin, C.J.}, \bibinfo{author}{Morales, N.},
  \bibinfo{author}{Morin, J.}, \bibinfo{author}{Nelson, R.P.},
  \bibinfo{author}{Ortiz, J.}, \bibinfo{author}{Ofir, A.},
  \bibinfo{author}{Paardekooper, S.J.}, \bibinfo{author}{Reiners, A.},
  \bibinfo{author}{Rodr{\'\i}guez, E.}, \bibinfo{author}{Rodrίguez-L{\'o}pez,
  C.}, \bibinfo{author}{Sarmiento, L.F.}, \bibinfo{author}{Strachan, J.P.},
  \bibinfo{author}{Tsapras, Y.}, \bibinfo{author}{Tuomi, M.},
  \bibinfo{author}{Zechmeister, M.}, \bibinfo{year}{2016}.
\newblock \bibinfo{title}{A terrestrial planet candidate in a temperate orbit
  around proxima centauri}.
\newblock \bibinfo{journal}{Nature} \bibinfo{volume}{536},
  \bibinfo{pages}{437--440}.
\newblock \URLprefix \url{http://dx.doi.org/10.1038/nature19106}.
\bibitem[{Anglada-Escud{\'e} and Tuomi(2015)}]{Anglada2015}
\bibinfo{author}{Anglada-Escud{\'e}, G.}, \bibinfo{author}{Tuomi, M.},
  \bibinfo{year}{2015}.
\newblock \bibinfo{title}{Comment on “stellar activity masquerading as
  planets in the habitable zone of the m dwarf gliese 581”}.
\newblock \bibinfo{journal}{Science} \bibinfo{volume}{347},
  \bibinfo{pages}{1080--1080}.
\bibitem[{{Aubert} et~al.(2009){Aubert}, {Labrosse} and {Poitou}}]{Aubert09}
\bibinfo{author}{{Aubert}, J.}, \bibinfo{author}{{Labrosse}, S.},
  \bibinfo{author}{{Poitou}, C.}, \bibinfo{year}{2009}.
\newblock \bibinfo{title}{{Modelling the palaeo-evolution of the geodynamo}}.
\newblock \bibinfo{journal}{Geophysical Journal International}
  \bibinfo{volume}{179}, \bibinfo{pages}{1414--1428}.
\newblock \DOIprefix\doi{10.1111/j.1365-246X.2009.04361.x}.
\bibitem[{{Barnes} et~al.(2016){Barnes}, {Deitrick}, {Luger}, {Driscoll},
  {Quinn}, {Fleming}, {Guyer}, {McDonald}, {Meadows}, {Arney}, {Crisp},
  {Domagal-Goldman}, {Lincowski}, {Lustig-Yaeger} and
  {Schwieterman}}]{Barnes2016}
\bibinfo{author}{{Barnes}, R.}, \bibinfo{author}{{Deitrick}, R.},
  \bibinfo{author}{{Luger}, R.}, \bibinfo{author}{{Driscoll}, P.E.},
  \bibinfo{author}{{Quinn}, T.R.}, \bibinfo{author}{{Fleming}, D.P.},
  \bibinfo{author}{{Guyer}, B.}, \bibinfo{author}{{McDonald}, D.V.},
  \bibinfo{author}{{Meadows}, V.S.}, \bibinfo{author}{{Arney}, G.},
  \bibinfo{author}{{Crisp}, D.}, \bibinfo{author}{{Domagal-Goldman}, S.D.},
  \bibinfo{author}{{Lincowski}, A.}, \bibinfo{author}{{Lustig-Yaeger}, J.},
  \bibinfo{author}{{Schwieterman}, E.}, \bibinfo{year}{2016}.
\newblock \bibinfo{title}{{The Habitability of Proxima Centauri b I:
  Evolutionary Scenarios}}.
\newblock \bibinfo{journal}{ArXiv e-prints}
  \href{http://arxiv.org/abs/1608.06919}{\tt arXiv:1608.06919}.
\bibitem[{Bazot et~al.(2016)Bazot, Christensen-Dalsgaard, Gizon and
  Benomar}]{Bazot2016}
\bibinfo{author}{Bazot, M.}, \bibinfo{author}{Christensen-Dalsgaard, J.},
  \bibinfo{author}{Gizon, L.}, \bibinfo{author}{Benomar, O.},
  \bibinfo{year}{2016}.
\newblock \bibinfo{title}{On the uncertain nature of the core of $\alpha$ cen
  a}.
\newblock \bibinfo{journal}{Monthly Notices of the Royal Astronomical Society}
  \bibinfo{volume}{460}, \bibinfo{pages}{1254--1269}.
\bibitem[{Chen and Kipping(2016)}]{Chen2016}
\bibinfo{author}{Chen, J.}, \bibinfo{author}{Kipping, D.M.},
  \bibinfo{year}{2016}.
\newblock \bibinfo{title}{Probabilistic forecasting of the masses and radii of
  other worlds}.
\newblock \bibinfo{journal}{arXiv preprint arXiv:1603.08614} .
\bibitem[{Chill{\`a} and Schumacher(2012)}]{Chilla2012}
\bibinfo{author}{Chill{\`a}, F.}, \bibinfo{author}{Schumacher, J.},
  \bibinfo{year}{2012}.
\newblock \bibinfo{title}{New perspectives in turbulent rayleigh-b{\'e}nard
  convection}.
\newblock \bibinfo{journal}{The European Physical Journal E: Soft Matter and
  Biological Physics} \bibinfo{volume}{35}, \bibinfo{pages}{1--25}.
\bibitem[{{Christensen}(2010)}]{Christensen10}
\bibinfo{author}{{Christensen}, U.R.}, \bibinfo{year}{2010}.
\newblock \bibinfo{title}{{Dynamo Scaling Laws and Applications to the
  Planets}}.
\newblock \bibinfo{journal}{Space science reviews} \bibinfo{volume}{152},
  \bibinfo{pages}{565--590}.
\newblock \DOIprefix\doi{10.1007/s11214-009-9553-2}.
\bibitem[{{Coleman} et~al.(2016){Coleman}, {Nelson}, {Paardekooper},
  {Dreizler}, {Giesers} and {Anglada-Escude}}]{Coleman2016}
\bibinfo{author}{{Coleman}, G.A.L.}, \bibinfo{author}{{Nelson}, R.P.},
  \bibinfo{author}{{Paardekooper}, S.J.}, \bibinfo{author}{{Dreizler}, S.},
  \bibinfo{author}{{Giesers}, B.}, \bibinfo{author}{{Anglada-Escude}, G.},
  \bibinfo{year}{2016}.
\newblock \bibinfo{title}{{Exploring plausible formation scenarios for the
  planet candidate orbiting Proxima Centauri}}.
\newblock \bibinfo{journal}{ArXiv e-prints}
  \href{http://arxiv.org/abs/1608.06908}{\tt arXiv:1608.06908}.
\bibitem[{Cuartas-Restrepo et~al.(2016)Cuartas-Restrepo, Melita, Zuluaga,
  Portilla, Sucerquia and Miloni}]{Cuartas2016}
\bibinfo{author}{Cuartas-Restrepo, P.}, \bibinfo{author}{Melita, M.},
  \bibinfo{author}{Zuluaga, J.}, \bibinfo{author}{Portilla, B.},
  \bibinfo{author}{Sucerquia, M.}, \bibinfo{author}{Miloni, O.},
  \bibinfo{year}{2016}.
\newblock \bibinfo{title}{The spin-orbit evolution of gj 667c system: The
  effect of composition and other planet's perturbations}.
\newblock \bibinfo{journal}{arXiv preprint arXiv:1606.07546} .
\bibitem[{Driscoll and Barnes(2015)}]{Driscoll2015}
\bibinfo{author}{Driscoll, P.}, \bibinfo{author}{Barnes, R.},
  \bibinfo{year}{2015}.
\newblock \bibinfo{title}{Tidal heating of earth-like exoplanets around m
  stars: thermal, magnetic, and orbital evolutions}.
\newblock \bibinfo{journal}{Astrobiology} \bibinfo{volume}{15},
  \bibinfo{pages}{739--760}.
\bibitem[{Dziewonski and Anderson(1981)}]{Dziewonski1981}
\bibinfo{author}{Dziewonski, A.M.}, \bibinfo{author}{Anderson, D.L.},
  \bibinfo{year}{1981}.
\newblock \bibinfo{title}{Preliminary reference earth model}.
\newblock \bibinfo{journal}{Physics of the earth and planetary interiors}
  \bibinfo{volume}{25}, \bibinfo{pages}{297--356}.
\bibitem[{{Gaidos} et~al.(2010){Gaidos}, {Conrad}, {Manga} and
  {Hernlund}}]{Gaidos10}
\bibinfo{author}{{Gaidos}, E.}, \bibinfo{author}{{Conrad}, C.P.},
  \bibinfo{author}{{Manga}, M.}, \bibinfo{author}{{Hernlund}, J.},
  \bibinfo{year}{2010}.
\newblock \bibinfo{title}{{Thermodynamic Limits on Magnetodynamos in Rocky
  Exoplanets}}.
\newblock \bibinfo{journal}{\apj} \bibinfo{volume}{718},
  \bibinfo{pages}{596--609}.
\newblock \DOIprefix\doi{10.1088/0004-637X/718/2/596},
  \href{http://arxiv.org/abs/1005.3523}{\tt arXiv:1005.3523}.
\bibitem[{Gastine et~al.(2016)Gastine, Wicht and Aubert}]{Gastine2016}
\bibinfo{author}{Gastine, T.}, \bibinfo{author}{Wicht, J.},
  \bibinfo{author}{Aubert, J.}, \bibinfo{year}{2016}.
\newblock \bibinfo{title}{Scaling regimes in spherical shell rotating
  convection}.
\newblock \bibinfo{journal}{Journal of Fluid Mechanics} \bibinfo{volume}{808},
  \bibinfo{pages}{690--732}.
\bibitem[{Gomi et~al.(2013)Gomi, Ohta, Hirose, Labrosse, Caracas, Verstraete
  and Hernlund}]{Gomi2013}
\bibinfo{author}{Gomi, H.}, \bibinfo{author}{Ohta, K.},
  \bibinfo{author}{Hirose, K.}, \bibinfo{author}{Labrosse, S.},
  \bibinfo{author}{Caracas, R.}, \bibinfo{author}{Verstraete, M.J.},
  \bibinfo{author}{Hernlund, J.W.}, \bibinfo{year}{2013}.
\newblock \bibinfo{title}{The high conductivity of iron and thermal evolution
  of the earth’s core}.
\newblock \bibinfo{journal}{Physics of the Earth and Planetary Interiors}
  \bibinfo{volume}{224}, \bibinfo{pages}{88--103}.
\bibitem[{Harder and Schubert(2001)}]{Harder2001}
\bibinfo{author}{Harder, H.}, \bibinfo{author}{Schubert, G.},
  \bibinfo{year}{2001}.
\newblock \bibinfo{title}{Sulfur in mercury's core?}
\newblock \bibinfo{journal}{Icarus} \bibinfo{volume}{151},
  \bibinfo{pages}{118--122}.
\bibitem[{Hatzes(2016)}]{Hatzes2016}
\bibinfo{author}{Hatzes, A.P.}, \bibinfo{year}{2016}.
\newblock \bibinfo{title}{Periodic h$\alpha$ variations in gl 581: Further
  evidence for an activity origin to gl 581d}.
\newblock \bibinfo{journal}{Astronomy \& Astrophysics} \bibinfo{volume}{585},
  \bibinfo{pages}{A144}.
\bibitem[{Hinkel and Kane(2013)}]{Hinkel2013}
\bibinfo{author}{Hinkel, N.R.}, \bibinfo{author}{Kane, S.R.},
  \bibinfo{year}{2013}.
\newblock \bibinfo{title}{Implications of the spectroscopic abundances in
  $\alpha$ centauri a and b}.
\newblock \bibinfo{journal}{Monthly Notices of the Royal Astronomical Society:
  Letters} , \bibinfo{pages}{slt032}.
\bibitem[{Hunter et~al.(2007)}]{Hunter2007}
\bibinfo{author}{Hunter, J.D.}, et~al., \bibinfo{year}{2007}.
\newblock \bibinfo{title}{Matplotlib: A 2d graphics environment}.
\newblock \bibinfo{journal}{Computing in science and engineering}
  \bibinfo{volume}{9}, \bibinfo{pages}{90--95}.
\bibitem[{{KamLAND Collaboration} et~al.(2011)}]{Kamland2011}
\bibinfo{author}{{KamLAND Collaboration}}, et~al., \bibinfo{year}{2011}.
\newblock \bibinfo{title}{Partial radiogenic heat model for earth revealed by
  geoneutrino measurements}.
\newblock \bibinfo{journal}{Nature Geoscience} \bibinfo{volume}{4},
  \bibinfo{pages}{647--651}.
\bibitem[{Kipping et~al.(2016)Kipping, Cameron, Hartman, Davenport, Matthews,
  Sasselov, Rowe, Siverd, Chen, Sandford et~al.}]{Kipping2016}
\bibinfo{author}{Kipping, D.M.}, \bibinfo{author}{Cameron, C.},
  \bibinfo{author}{Hartman, J.D.}, \bibinfo{author}{Davenport, J.R.},
  \bibinfo{author}{Matthews, J.M.}, \bibinfo{author}{Sasselov, D.},
  \bibinfo{author}{Rowe, J.}, \bibinfo{author}{Siverd, R.J.},
  \bibinfo{author}{Chen, J.}, \bibinfo{author}{Sandford, E.}, et~al.,
  \bibinfo{year}{2016}.
\newblock \bibinfo{title}{No conclusive evidence for transits of proxima b in
  most photometry}.
\newblock \bibinfo{journal}{arXiv preprint arXiv:1609.08718} .
\bibitem[{{Kite} et~al.(2009){Kite}, {Manga} and {Gaidos}}]{Kite09}
\bibinfo{author}{{Kite}, E.S.}, \bibinfo{author}{{Manga}, M.},
  \bibinfo{author}{{Gaidos}, E.}, \bibinfo{year}{2009}.
\newblock \bibinfo{title}{{Geodynamics and Rate of Volcanism on Massive
  Earth-like Planets}}.
\newblock \bibinfo{journal}{\apj} \bibinfo{volume}{700},
  \bibinfo{pages}{1732--1749}.
\newblock \DOIprefix\doi{10.1088/0004-637X/700/2/1732},
  \href{http://arxiv.org/abs/0809.2305}{\tt arXiv:0809.2305}.
\bibitem[{{Labrosse}(2003)}]{Labrosse03}
\bibinfo{author}{{Labrosse}, S.}, \bibinfo{year}{2003}.
\newblock \bibinfo{title}{{Thermal and magnetic evolution of the Earth's
  core}}.
\newblock \bibinfo{journal}{Physics of the Earth and Planetary Interiors}
  \bibinfo{volume}{140}, \bibinfo{pages}{127--143}.
\newblock \DOIprefix\doi{10.1016/j.pepi.2003.07.006}.
\bibitem[{{Labrosse} et~al.(2001){Labrosse}, {Poirier} and {Le
  Mou{\"e}l}}]{Labrosse01}
\bibinfo{author}{{Labrosse}, S.}, \bibinfo{author}{{Poirier}, J.P.},
  \bibinfo{author}{{Le Mou{\"e}l}, J.L.}, \bibinfo{year}{2001}.
\newblock \bibinfo{title}{{The age of the inner core}}.
\newblock \bibinfo{journal}{Earth and Planetary Science Letters}
  \bibinfo{volume}{190}, \bibinfo{pages}{111--123}.
\newblock \DOIprefix\doi{10.1016/S0012-821X(01)00387-9}.
\bibitem[{Lau et~al.(2016)Lau, Mitrovica, Austermann, Crawford, Al-Attar and
  Latychev}]{Lau2016}
\bibinfo{author}{Lau, H.C.}, \bibinfo{author}{Mitrovica, J.X.},
  \bibinfo{author}{Austermann, J.}, \bibinfo{author}{Crawford, O.},
  \bibinfo{author}{Al-Attar, D.}, \bibinfo{author}{Latychev, K.},
  \bibinfo{year}{2016}.
\newblock \bibinfo{title}{Inferences of mantle viscosity based on ice age data
  sets: Radial structure}.
\newblock \bibinfo{journal}{Journal of Geophysical Research: Solid Earth}
  \bibinfo{volume}{121}, \bibinfo{pages}{6991--7012}.
\bibitem[{L{\'e}ger et~al.(2004)L{\'e}ger, Selsis, Sotin, Guillot, Despois,
  Mawet, Ollivier, Lab{\`e}que, Valette, Brachet et~al.}]{Leger2004}
\bibinfo{author}{L{\'e}ger, A.}, \bibinfo{author}{Selsis, F.},
  \bibinfo{author}{Sotin, C.}, \bibinfo{author}{Guillot, T.},
  \bibinfo{author}{Despois, D.}, \bibinfo{author}{Mawet, D.},
  \bibinfo{author}{Ollivier, M.}, \bibinfo{author}{Lab{\`e}que, A.},
  \bibinfo{author}{Valette, C.}, \bibinfo{author}{Brachet, F.}, et~al.,
  \bibinfo{year}{2004}.
\newblock \bibinfo{title}{A new family of planets?“ocean-planets”}.
\newblock \bibinfo{journal}{Icarus} \bibinfo{volume}{169},
  \bibinfo{pages}{499--504}.
\bibitem[{Lubin(2016)}]{Lubin2016}
\bibinfo{author}{Lubin, P.}, \bibinfo{year}{2016}.
\newblock \bibinfo{title}{A roadmap to interstellar flight}.
\newblock \bibinfo{journal}{arXiv preprint arXiv:1604.01356} .
\bibitem[{Makarov(2013)}]{Makarov2013a}
\bibinfo{author}{Makarov, V.V.}, \bibinfo{year}{2013}.
\newblock \bibinfo{title}{Why is the moon synchronously rotating?}
\newblock \bibinfo{journal}{Monthly Notices of the Royal Astronomical Society:
  Letters} , \bibinfo{pages}{slt068}.
\bibitem[{Makarov and Efroimsky(2013)}]{Makarov2013b}
\bibinfo{author}{Makarov, V.V.}, \bibinfo{author}{Efroimsky, M.},
  \bibinfo{year}{2013}.
\newblock \bibinfo{title}{No pseudosynchronous rotation for terrestrial planets
  and moons}.
\newblock \bibinfo{journal}{The Astrophysical Journal} \bibinfo{volume}{764},
  \bibinfo{pages}{27}.
\bibitem[{Malkus(1954a)}]{Malkus54b}
\bibinfo{author}{Malkus, M.}, \bibinfo{year}{1954}a.
\newblock \bibinfo{title}{Discrete transitions in turbulent convection}.
\newblock \bibinfo{journal}{Proceedings of the Royal Society of London A:
  Mathematical, Physical and Engineering Sciences} \bibinfo{volume}{225},
  \bibinfo{pages}{185--195}.
\newblock \URLprefix
  \url{http://rspa.royalsocietypublishing.org/content/225/1161/185},
  \DOIprefix\doi{10.1098/rspa.1954.0196},
  \href{http://arxiv.org/abs/http://rspa.royalsocietypublishing.org/content/225/1161/185.full.pdf}{\tt
  arXiv:http://rspa.royalsocietypublishing.org/content/225/1161/185.full.pdf}.
\bibitem[{Malkus(1954b)}]{Malkus54a}
\bibinfo{author}{Malkus, M.}, \bibinfo{year}{1954}b.
\newblock \bibinfo{title}{The heat transport and spectrum of thermal
  turbulence}.
\newblock \bibinfo{journal}{Proceedings of the Royal Society of London A:
  Mathematical, Physical and Engineering Sciences} \bibinfo{volume}{225},
  \bibinfo{pages}{196--212}.
\newblock \URLprefix
  \url{http://rspa.royalsocietypublishing.org/content/225/1161/196},
  \DOIprefix\doi{10.1098/rspa.1954.0197},
  \href{http://arxiv.org/abs/http://rspa.royalsocietypublishing.org/content/225/1161/196.full.pdf}{\tt
  arXiv:http://rspa.royalsocietypublishing.org/content/225/1161/196.full.pdf}.
\bibitem[{Mitrovica and Forte(2004)}]{Mitrovica2004}
\bibinfo{author}{Mitrovica, J.}, \bibinfo{author}{Forte, A.},
  \bibinfo{year}{2004}.
\newblock \bibinfo{title}{A new inference of mantle viscosity based upon joint
  inversion of convection and glacial isostatic adjustment data}.
\newblock \bibinfo{journal}{Earth and Planetary Science Letters}
  \bibinfo{volume}{225}, \bibinfo{pages}{177--189}.
\bibitem[{{Nimmo}(2009a)}]{Nimmo09b}
\bibinfo{author}{{Nimmo}, F.}, \bibinfo{year}{2009}a.
\newblock \bibinfo{title}{{Energetics of asteroid dynamos and the role of
  compositional convection}}.
\newblock \bibinfo{journal}{Geophys. Res. Lett.} \bibinfo{volume}{36},
  \bibinfo{pages}{L10201}.
\newblock \DOIprefix\doi{10.1029/2009GL037997}.
\bibitem[{{Nimmo}(2009b)}]{Nimmo09a}
\bibinfo{author}{{Nimmo}, F.}, \bibinfo{year}{2009}b.
\newblock \bibinfo{title}{Treatise on Geophysics}.
  \bibinfo{publisher}{Elsevier}. volume~\bibinfo{volume}{8}. chapter
  \bibinfo{chapter}{Enegetics of the Core}.
\newblock pp. \bibinfo{pages}{31--68}.
\bibitem[{P{\'e}rez and Granger(2007)}]{Perez2007}
\bibinfo{author}{P{\'e}rez, F.}, \bibinfo{author}{Granger, B.E.},
  \bibinfo{year}{2007}.
\newblock \bibinfo{title}{Ipython: a system for interactive scientific
  computing}.
\newblock \bibinfo{journal}{Computing in Science \& Engineering}
  \bibinfo{volume}{9}, \bibinfo{pages}{21--29}.
\bibitem[{{Pozzo} et~al.(2012){Pozzo}, {Davies}, {Gubbins} and
  {Alf{\`e}}}]{Pozzo12}
\bibinfo{author}{{Pozzo}, M.}, \bibinfo{author}{{Davies}, C.},
  \bibinfo{author}{{Gubbins}, D.}, \bibinfo{author}{{Alf{\`e}}, D.},
  \bibinfo{year}{2012}.
\newblock \bibinfo{title}{{Thermal and electrical conductivity of iron at
  Earth's core conditions}}.
\newblock \bibinfo{journal}{Nature} \bibinfo{volume}{485},
  \bibinfo{pages}{355--358}.
\newblock \DOIprefix\doi{10.1038/nature11031},
  \href{http://arxiv.org/abs/1203.4970}{\tt arXiv:1203.4970}.
\bibitem[{{Ribas} et~al.(2016){Ribas}, {Bolmont}, {Selsis}, {Reiners},
  {Leconte}, {Raymond}, {Engle}, {Guinan}, {Morin}, {Turbet}, {Forget} and
  {Anglada-Escude}}]{Ribas2016}
\bibinfo{author}{{Ribas}, I.}, \bibinfo{author}{{Bolmont}, E.},
  \bibinfo{author}{{Selsis}, F.}, \bibinfo{author}{{Reiners}, A.},
  \bibinfo{author}{{Leconte}, J.}, \bibinfo{author}{{Raymond}, S.N.},
  \bibinfo{author}{{Engle}, S.G.}, \bibinfo{author}{{Guinan}, E.F.},
  \bibinfo{author}{{Morin}, J.}, \bibinfo{author}{{Turbet}, M.},
  \bibinfo{author}{{Forget}, F.}, \bibinfo{author}{{Anglada-Escude}, G.},
  \bibinfo{year}{2016}.
\newblock \bibinfo{title}{{The habitability of Proxima Centauri b. I.
  Irradiation, rotation and volatile inventory from formation to the present}}.
\newblock \bibinfo{journal}{ArXiv e-prints}
  \href{http://arxiv.org/abs/1608.06813}{\tt arXiv:1608.06813}.
\bibitem[{{Ricard}(2009)}]{Ricard09}
\bibinfo{author}{{Ricard}, Y.}, \bibinfo{year}{2009}.
\newblock \bibinfo{title}{Treatise on Geophysics}.
  \bibinfo{publisher}{Elsevier}. volume~\bibinfo{volume}{7}. chapter
  \bibinfo{chapter}{Physics of Mantle Convection}.
\newblock p. \bibinfo{pages}{6054}.
\bibitem[{Stacey and Davis(2004)}]{Stacey2004}
\bibinfo{author}{Stacey, F.}, \bibinfo{author}{Davis, P.},
  \bibinfo{year}{2004}.
\newblock \bibinfo{title}{High pressure equations of state with applications to
  the lower mantle and core}.
\newblock \bibinfo{journal}{Physics of the Earth and Planetary Interiors}
  \bibinfo{volume}{142}, \bibinfo{pages}{137--184}.
\bibitem[{{Stamenkovi{\'c}} et~al.(2011){Stamenkovi{\'c}}, {Breuer} and
  {Spohn}}]{Stamenkovic11}
\bibinfo{author}{{Stamenkovi{\'c}}, V.}, \bibinfo{author}{{Breuer}, D.},
  \bibinfo{author}{{Spohn}, T.}, \bibinfo{year}{2011}.
\newblock \bibinfo{title}{{Thermal and transport properties of mantle rock at
  high pressure: Applications to super-Earths}}.
\newblock \bibinfo{journal}{Icarus} \bibinfo{volume}{216},
  \bibinfo{pages}{572--596}.
\newblock \DOIprefix\doi{10.1016/j.icarus.2011.09.030}.
\bibitem[{{Tachinami} et~al.(2011){Tachinami}, {Senshu} and
  {Ida}}]{Tachinami11}
\bibinfo{author}{{Tachinami}, C.}, \bibinfo{author}{{Senshu}, H.},
  \bibinfo{author}{{Ida}, S.}, \bibinfo{year}{2011}.
\newblock \bibinfo{title}{{Thermal Evolution and Lifetime of Intrinsic Magnetic
  Fields of Super-Earths in Habitable Zones}}.
\newblock \bibinfo{journal}{\apj} \bibinfo{volume}{726},
  \bibinfo{pages}{70--87}.
\newblock \DOIprefix\doi{10.1088/0004-637X/726/2/70},
  \href{http://arxiv.org/abs/1010.5317}{\tt arXiv:1010.5317}.
\bibitem[{Th{\'e}venin et~al.(2002)Th{\'e}venin, Provost, Morel, Berthomieu,
  Bouchy and Carrier}]{Thevenin2002}
\bibinfo{author}{Th{\'e}venin, F.}, \bibinfo{author}{Provost, J.},
  \bibinfo{author}{Morel, P.}, \bibinfo{author}{Berthomieu, G.},
  \bibinfo{author}{Bouchy, F.}, \bibinfo{author}{Carrier, F.},
  \bibinfo{year}{2002}.
\newblock \bibinfo{title}{Asteroseismology and calibration of $\alpha$ cen
  binary system}.
\newblock \bibinfo{journal}{Astronomy \& Astrophysics} \bibinfo{volume}{392},
  \bibinfo{pages}{L9--L12}.
\bibitem[{Thoul et~al.(2003)Thoul, Scuflaire, Noels, Vatovez, Briquet, Dupret
  and Montalban}]{Thoul2003}
\bibinfo{author}{Thoul, A.}, \bibinfo{author}{Scuflaire, R.},
  \bibinfo{author}{Noels, A.}, \bibinfo{author}{Vatovez, B.},
  \bibinfo{author}{Briquet, M.}, \bibinfo{author}{Dupret, M.A.},
  \bibinfo{author}{Montalban, J.}, \bibinfo{year}{2003}.
\newblock \bibinfo{title}{A new seismic analysis of alpha centauri}.
\newblock \bibinfo{journal}{Astronomy \& Astrophysics} \bibinfo{volume}{402},
  \bibinfo{pages}{293--297}.
\bibitem[{{Turbet} et~al.(2016){Turbet}, {Leconte}, {Selsis}, {Bolmont},
  {Forget}, {Ribas}, {Raymond} and {Anglada-Escud{\'e}}}]{Turbet2016}
\bibinfo{author}{{Turbet}, M.}, \bibinfo{author}{{Leconte}, J.},
  \bibinfo{author}{{Selsis}, F.}, \bibinfo{author}{{Bolmont}, E.},
  \bibinfo{author}{{Forget}, F.}, \bibinfo{author}{{Ribas}, I.},
  \bibinfo{author}{{Raymond}, S.N.}, \bibinfo{author}{{Anglada-Escud{\'e}},
  G.}, \bibinfo{year}{2016}.
\newblock \bibinfo{title}{{The habitability of Proxima Centauri b II. Possible
  climates and Observability}}.
\newblock \bibinfo{journal}{ArXiv e-prints}
  \href{http://arxiv.org/abs/1608.06827}{\tt arXiv:1608.06827}.
\bibitem[{{Valencia} et~al.(2006){Valencia}, {O'Connell} and
  {Sasselov}}]{Valencia06}
\bibinfo{author}{{Valencia}, D.}, \bibinfo{author}{{O'Connell}, R.J.},
  \bibinfo{author}{{Sasselov}, D.}, \bibinfo{year}{2006}.
\newblock \bibinfo{title}{{Internal structure of massive terrestrial planets}}.
\newblock \bibinfo{journal}{Icarus} \bibinfo{volume}{181},
  \bibinfo{pages}{545--554}.
\newblock \DOIprefix\doi{10.1016/j.icarus.2005.11.021},
  \href{http://arxiv.org/abs/arXiv:astro-ph/0511150}{\tt
  arXiv:arXiv:astro-ph/0511150}.
\bibitem[{{Valencia} et~al.(2007a){Valencia}, {Sasselov} and
  {O'Connell}}]{Valencia07a}
\bibinfo{author}{{Valencia}, D.}, \bibinfo{author}{{Sasselov}, D.D.},
  \bibinfo{author}{{O'Connell}, R.J.}, \bibinfo{year}{2007}a.
\newblock \bibinfo{title}{{Detailed Models of Super-Earths: How Well Can We
  Infer Bulk Properties?}}
\newblock \bibinfo{journal}{\apj} \bibinfo{volume}{665},
  \bibinfo{pages}{1413--1420}.
\newblock \DOIprefix\doi{10.1086/519554},
  \href{http://arxiv.org/abs/0704.3454}{\tt arXiv:0704.3454}.
\bibitem[{{Valencia} et~al.(2007b){Valencia}, {Sasselov} and
  {O'Connell}}]{Valencia07b}
\bibinfo{author}{{Valencia}, D.}, \bibinfo{author}{{Sasselov}, D.D.},
  \bibinfo{author}{{O'Connell}, R.J.}, \bibinfo{year}{2007}b.
\newblock \bibinfo{title}{{Radius and Structure Models of the First Super-Earth
  Planet}}.
\newblock \bibinfo{journal}{\apj} \bibinfo{volume}{656},
  \bibinfo{pages}{545--551}.
\newblock \DOIprefix\doi{10.1086/509800},
  \href{http://arxiv.org/abs/arXiv:astro-ph/0610122}{\tt
  arXiv:arXiv:astro-ph/0610122}.
\bibitem[{Van Der~Walt et~al.(2011)Van Der~Walt, Colbert and
  Varoquaux}]{Van2011}
\bibinfo{author}{Van Der~Walt, S.}, \bibinfo{author}{Colbert, S.C.},
  \bibinfo{author}{Varoquaux, G.}, \bibinfo{year}{2011}.
\newblock \bibinfo{title}{The numpy array: a structure for efficient numerical
  computation}.
\newblock \bibinfo{journal}{Computing in Science \& Engineering}
  \bibinfo{volume}{13}, \bibinfo{pages}{22--30}.
\bibitem[{Wolanin et~al.(1997)Wolanin, Pruzan, Chervin, Canny, Gauthier,
  H{\"a}usermann and Hanfland}]{Wolanin1997}
\bibinfo{author}{Wolanin, E.}, \bibinfo{author}{Pruzan, P.},
  \bibinfo{author}{Chervin, J.}, \bibinfo{author}{Canny, B.},
  \bibinfo{author}{Gauthier, M.}, \bibinfo{author}{H{\"a}usermann, D.},
  \bibinfo{author}{Hanfland, M.}, \bibinfo{year}{1997}.
\newblock \bibinfo{title}{Equation of state of ice vii up to 106 gpa}.
\newblock \bibinfo{journal}{Physical Review B} \bibinfo{volume}{56},
  \bibinfo{pages}{5781}.
\bibitem[{{Yamazaki} and {Karato}(2001)}]{Yamazaki01}
\bibinfo{author}{{Yamazaki}, D.}, \bibinfo{author}{{Karato}, S.},
  \bibinfo{year}{2001}.
\newblock \bibinfo{title}{{Some mineral physics constraints on the rheology and
  geothermal structure of Earth???s lower mantle}}.
\newblock \bibinfo{journal}{American Mineralogist} \bibinfo{volume}{86},
  \bibinfo{pages}{385--391}.
\bibitem[{{Zuluaga} et~al.(2013){Zuluaga}, {Bustamante}, {Cuartas} and
  {Hoyos}}]{Zuluaga2013}
\bibinfo{author}{{Zuluaga}, J.I.}, \bibinfo{author}{{Bustamante}, S.},
  \bibinfo{author}{{Cuartas}, P.A.}, \bibinfo{author}{{Hoyos}, J.H.},
  \bibinfo{year}{2013}.
\newblock \bibinfo{title}{{The Influence of Thermal Evolution in the Magnetic
  Protection of Terrestrial Planets}}.
\newblock \bibinfo{journal}{\apj} \bibinfo{volume}{770}, \bibinfo{pages}{23}.
\newblock \DOIprefix\doi{10.1088/0004-637X/770/1/23},
  \href{http://arxiv.org/abs/1304.2909}{\tt arXiv:1304.2909}.
\bibitem[{{Zuluaga} and {Cuartas}(2012)}]{Zuluaga12}
\bibinfo{author}{{Zuluaga}, J.I.}, \bibinfo{author}{{Cuartas}, P.A.},
  \bibinfo{year}{2012}.
\newblock \bibinfo{title}{{The role of rotation in the evolution of
  dynamo-generated magnetic fields in Super Earths}}.
\newblock \bibinfo{journal}{Icarus} \bibinfo{volume}{217},
  \bibinfo{pages}{88--102}.
\newblock \DOIprefix\doi{10.1016/j.icarus.2011.10.014},
  \href{http://arxiv.org/abs/1101.0691}{\tt arXiv:1101.0691}.

\end{thebibliography}
\end{document}